\documentstyle[12pt,a4]{article}
\title{\sf SHELL EFFECTS IN NUCLEI WITH VECTOR SELF-COUPLING OF OMEGA
MESON IN RELATIVISTIC HARTREE-BOGOLIUBOV THEORY}
\author{Madan M. Sharma, Ameenah R. Farhan and S. Mythili\\
Physics Department, Kuwait University, Kuwait 13060}

\begin{document}
 
\maketitle
\begin{abstract}

Shell effects in nuclei about the stability line are investigated
within the framework of the Relativistic Hartree-Bogoliubov (RHB) 
theory with self-consistent finite-range pairing. Using 2-neutron 
separation energies of Ni and Sn isotopes, the role of $\sigma$- and
$\omega$-meson couplings on the shell effects in nuclei is examined.
It is observed that the existing successful nuclear forces 
(Lagrangian parameter sets) based upon the nonlinear scalar coupling 
of $\sigma$-meson exhibit shell effects which are stronger than 
suggested by the experimental data. We have introduced nonlinear 
vector self-coupling of $\omega$-meson in the RHB theory. It is shown 
that the inclusion of the vector self-coupling of $\omega$-meson 
in addition to the nonlinear scalar coupling of $\sigma$-meson 
provides a good agreement with the experimental data on shell effects 
in nuclei about the stability line. A comparison of the shell effects 
in the RHB theory is made with the Hartree-Fock Bogoliubov approach 
using the Skyrme force SkP. It is shown that the oft-discussed
shell quenching with SkP is not consistent with the available 
experimental data. 

\end{abstract}

\newpage
\baselineskip=24pt

\section{\bf Introduction}

The Relativistic Mean-Field (RMF) theory \cite{SW.86} has proved to be 
successful in providing a framework for description of various facets of 
nuclear properties \cite{Rein.89,GRT.90,Ser.92,SNR.93,Ring.96,SW.98}.
In the RMF theory, the nuclear force is produced by a
virtual exchange of various mesons. The nuclear saturation is achieved 
by a balance between an attractive $\sigma$- and a repulsive 
$\omega$-field. The relativistic Lorentz covariance of the theory 
allows an intrinsic spin-orbit interaction based upon exchange of
$\sigma$- and $\omega$-mesons. This has been shown to be advantageous 
for properties which depend upon spin-orbit potential \cite{SLR.93}.
An immediate advantage of the proper spin-orbit potential has been the
success \cite{SLR.93} of the RMF theory to be able to describe the 
anomalous kink in the isotope shifts of Pb nuclei. The isotope shifts of 
Pb have been measured with a high precision using atomic beam 
laser spectroscopy \cite{Otten.89} and are known to show a 
pronounced kink about the magic number N=126. The non-relativistic 
approaches based upon the Skyrme and Gogny forces have been unable 
to reproduce this kink \cite{Tajima.93}. It was shown \cite{SLK.94} 
that this difference in the predictions of the Skyrme model and the RMF 
theory is due to the isospin dependence of the spin-orbit term.
The Skyrme model assumes an isospin independent two-body spin-orbit
force. A strong isospin dependence is, however, introduced herein
by the exchange term. On the other hand, the isospin dependence in
the RMF theory is provided mainly by the coupling constant $g_\rho$ of the 
$\rho$-meson. However, as the strength of the spin-orbit term derives
from a large sum of the absolute values of the scalar and vector fields,
the $\rho$-field is rendered much weaker in comparison. Consequently, 
the isospin dependence of the spin-orbit potential in the RMF theory 
is relatively weak. A weak isospin dependence of the spin-orbit
potential in the RMF theory was also concluded from calculations 
employing asymmetric semi-infinite nuclear matter \cite{Eiff.95}.

The RMF theory has achieved significant success in respect of nuclei
near the stability line as well as for nuclei far away from the
stability line \cite{GRT.90,SNR.93,Ring.96,LFS.98}.The binding energies,
charge radii, deformation properties and isotope shifts are some 
of the properties which are described successfully in the RMF theory.
The nonlinear scalar self-coupling of $\sigma$-meson  has been
the most successful model of the RMF theory used so far. With the
need to describe highly exotic nuclei far away from the stability line,
where the particles couple to the continuum, Relativistic 
Hartree-Bogoliubov (RHB) theory with a self-consistent pairing has
been developed \cite{Gon.96}. It has been shown \cite{Lala.98,Meng.98} that 
the RHB theory which uses the nonlinear scalar model provides a good 
description of nuclei away from the stability line. However, the level 
densities in the RMF or the RHB theory are perceived to be
generally low, which lead to larger shell gaps. The large shell gaps 
imply that shell effects are stronger than in the experimental data. 
In the present work, we focus upon the shell effects and investigate 
the role of various meson couplings on the shell effects.

The shell effects manifest strongly in nuclei with magic particle numbers. 
Such an effect is well known to exist experimentally in many nuclei all 
over the periodic table, whereby a stronger binding is exhibited by 
these nuclei as compared to neighbouring nuclei. Nuclei with a 
particle or two above a magic number show a significantly reduced
binding for the extraneous particles above the shell closure.
The origin of shell closure has long been understood \cite{MGM.55}
due to spin-orbit coupling and to an ensuing splitting of levels.
The spin-orbit potential and a bunching of levels create shell 
closures (magic numbers) which are predicted correctly by most of 
the models. In the non-relativistic density-dependent theory of the 
Skyrme type \cite{Vauth.73}, the spin-orbit interaction is added 
phenomenologically and its strength is adjusted to reproduce
the spin-orbit splitting in $^{16}$O. In the RMF theory, on the other
hand, the spin-orbit interaction arises naturally due to exchange of
$\sigma$ and $\omega$ mesons by the nucleons. The strength of the 
interaction is determined by the spin-orbit splitting in $^{16}$O
and other nuclei, which is uniquely decided by the effective mass.
This constrains an effective mass $m^*$ in the vicinity of $\sim
0.60$ in the RMF models \cite{Rein.89}.

The nature of the shell effects about the stability 
line is well established through the experimental  data 
available on the binding energies and on one- and two-particle 
separation energies over a large region of the periodic table.
The experimental data encompass all the known magic numbers.
Using the experimental binding energies on 2-neutron and 2-proton 
separation energies have been plotted in ref. \cite{Bor.93} for nuclei 
all over the periodic table. An unambiguous kink appears in the
separation energies across all the major magic numbers, 
demonstrating the existence of a major shell closure at these
numbers. This also shows that the shell effects at 
most of the major magic numbers along the line of stability
are strong. 

It is not clear how the magic numbers or the shell closures 
behave in going to extreme regions of the periodic table such as 
near the drip lines. Experimental data about neutron drip-line in the 
region of light nuclei are gradually emerging 
\cite{Rkl.92,MS.93,Tan.95,Ver.96,Proc.96}. 
As the neutron drip-line for medium-heavy and heavy nuclei is 
predicted to lie very far away (with abnormally large neutron 
excesses) from the stability line, it is not expected that experimental 
data on such nuclei would become available in the near future.
However, information on shell effects in such nuclei in the vicinity 
of drip lines are vital to understanding the r-process 
nucleosynthesis of heavy nuclei \cite{kratz.93}

Shell effects near the drip lines have been a matter of intense debate in 
the last few years. Within the framework of the RMF theory, it was
shown that shell effects about N=82 in the vicinity of the neutron drip-line 
remain strong \cite{SLH.94}. These conclusions were based upon the
RMF theory with the non-linear scalar potential of $\sigma$-meson.
On the other hand, on the basis of the Hartree-Fock-Bogoliubov approach 
with the Skyrme interaction SkP, it was contended that the shell
effects at N=82 in the vicinity of the neutron drip-line show a strong
quenching \cite{Doba.94}. However, in the absence of any data in this 
region in forseeable future, it is difficult to ascertain whether 
the shell effects near the drip-line remain strong or are quenched.
Therefore, this debate needs to looks into  other  observables or 
evidences which might support or refute the either contention. 
In this paper, we begin with our investigations of shell effects first
along the stability line where experimental data is available. Having
founded the basis, it could then be extended to predict behaviour of
the shell effects at the drip lines. With this in mind, we include
the nonlinear vector self-coupling of $\omega$-meson in the Lagrangian
of the RMF theory. 

The non-linear vector self-coupling of $\omega$-meson was introduced
by Bodmer \cite{Bod.91} and properties of nuclear matter were discussed 
on adding a quartic term in $\omega$-meson potential. The vector 
self-coupling to the properties of finite nuclei was applied in 
ref. \cite{Suga.94}. It was noticed \cite{Bod.91} that 
the inclusion of the vector self-coupling of $\omega$ meson in 
addition to the non-linear scalar self-coupling of $\sigma$ meson has 
the effect of softening the high-density equation of state (EOS) 
of the nuclear matter. The EOS of nuclear matter has been investigated 
\cite{Muller.96} by taking expansion terms of higher orders in the 
$\sigma$ and $\omega$ fields. Recently, a comprehensive study of the vector 
self-coupling and its effect on properties of nuclei and nuclear matter 
has been undertaken \cite{SH.99}. An emphasis has been given to various 
facets of finite nuclei. It is shown \cite{SH.99} that an inclusion 
of the vector self-coupling of $\omega$-meson gives an improved
description of nuclei as compared to the frequently used scalar 
self-coupling of $\sigma$-meson only. Two new forces NL-SV1 and
NL-SV2 have been proposed \cite{SH.99} with a view to improve 
the predictions of the ground-state properties of nuclei. We  will 
employ these forces in the present work. Details of this study will be
provided elsewhere \cite{SH.99}.

In this paper, we discuss the details of the RHB theory preceded by 
a short review of the RMF approach. The calculational details are
provided in the section thereafter. The main objective of this work 
is to investigate the shell effects in nuclei about the stability line
using the vector self-coupling of $\omega$ meson and compare
the results with those from the scalar self-coupling of $\sigma$ meson.
We have taken the isotopic chains of Ni and Sn nuclei for this
purpose. Experimental binding energies of these nuclei are available
over a large range. This allows us to examine
the shell effects in nuclei in the known region using the various
models. The shell effects about the magic neutron numbers N=28, 50 
and 82 are explored. The RMF+BCS results are compared with those
of the RHB theory. The results on  binding energies and
two-neutron separation energies are presented and discussed in
Section 5. The single-particle levels with various approaches are
presented for a few nuclei, with a view to reinforce the results
on the shell effects. The last section summarizes the main conclusions.

\section{\bf The Relativistic Mean-Field Theory}

The RMF approach \cite{SW.86} is based upon the Lagrangian density 
which consists of fields due to the various mesons interacting with 
the nucleons. The mesons include the isoscalar scalar $\sigma$-meson, 
the isovector vector $\omega$-meson and the isovector vector $\rho$-meson.
The Lagrangian density is given by:

\begin{equation}
\begin{array}{rl}
{\cal L} &=
\bar \psi (i\rlap{/}\partial -M) \psi +
\,{1\over2}\partial_\mu\sigma\partial^\mu\sigma-U(\sigma)
-{1\over4}\Omega_{\mu\nu}\Omega^{\mu\nu}+\\
\                                        \\
\ & {1\over2}m_\omega^2\omega_\mu\omega^\mu
+{1\over4}g_4(\omega_\mu\omega^\mu)^2
-{1\over4}{\bf R}_{\mu\nu}{\bf R}^{\mu\nu} +
 {1\over2}m_{\rho}^{2}
 \mbox{\boldmath $\rho$}_{\mu}\mbox{\boldmath $\rho$}^{\mu}
-{1\over4}F_{\mu\nu}F^{\mu\nu} \\
\                              \\
\ &  g_{\sigma}\bar\psi \sigma \psi~
     -~g_{\omega}\bar\psi \rlap{/}\omega \psi~
     -~g_{\rho}  \bar\psi
      \rlap{/}\mbox{\boldmath $\rho$}

     \mbox{\boldmath $\tau$} \psi
     -~e \bar\psi \rlap{/}{\bf A} \psi
\end{array}
\end{equation}
The bold-faced letters indicate the vector quantities. 
Here M, m$_{\sigma}$, m$_{\omega}$ and m$_{\rho}$ denote the nucleon-, 
the $\sigma$-, the $\omega$- and the $\rho$-meson masses respectively, 
while g$_{\sigma}$, g$_{\omega}$, g$_{\rho}$ and e$^2$/4$\pi$ = 1/137 are 
the corresponding coupling constants for the mesons and the photon,
respectively. 

The $\sigma$ meson is assumed to move in a scalar potential of the form:
\begin{equation}
U(\sigma)~={1\over2}m_{\sigma}^{2} \sigma^{2}~+~
{1\over3}g_{2}\sigma^{3}~+~{1\over4}g_{3}\sigma^{4}.
\end{equation}
This was introduced by Boguta and Bodmer \cite{BB.77} in order to
make a substantial improvement in the surface properties of finite
nuclei. This Ansatz for the $\sigma$-potential has since become a 
standard and necessary ingredient for description of the properties of 
finite nuclei. Recently, several variations of the non-linear $\sigma$ 
and $\omega$ fields have been proposed \cite{Furn.96}. 
The new proposals for the form of the Lagrangian need to be
investigated with a view to constrain the form in compliance with
the properties of nuclear matter and finite nuclei. 

Here we have incorporated the non-linear vector self-coupling of
the $\omega$-meson, in addition to the non-linear scalar potential
of Eq. (2). The coupling constant for the non-linear $\omega$-term
is denoted by $g_4$ in the Lagrangian (1). The vector self-coupling 
of the $\omega$-meson was first proposed in ref. \cite{Bod.91} where
properties of nuclear matter associated to this potential were
also discussed. However, the coupling constant of the new Ansatz 
can be constrained appropriately only within the framework of 
finite nuclei. One such attempt was made in ref. \cite{Suga.94} where  
parameter sets TM1 and TM2 were obtained. We have now performed  
a comprehensive investigation of the properties of nuclear matter 
and finite nuclei associated with the vector self-coupling 
\cite{SH.99}. 

The field tensors of the vector mesons and of the electromagnetic
field take the following form:
\begin{equation}
\begin {array}{rl}
\Omega^{\mu\nu} =& \partial^{\mu}\omega^{\nu}-\partial^{\nu}\omega^{\mu}\\
\          \\
{\bf R}^{\mu\nu} =& \partial^{\mu}
                  \mbox{\boldmath $\rho$}^{\nu}
                  -\partial^{\nu}
                  \mbox{\boldmath $\rho$}^{\mu}\\
\                \\
F^{\mu\nu} =& \partial^{\mu}{\bf A}^{\nu}-\partial^{\nu}{\bf A}^{\mu}
\end{array}
\end{equation}

The mean-field approximation constitutes the lowest order of the
quantum field theory. Herein, the nucleons are assumed to move
independently in the meson fields. The latter are replaced by their
classical expectation values. The ground-state of the nucleus is 
described by a Slater determinant $\vert\Phi >$ of single-particle
spinors $\psi_i$ (i = 1,2,....A). The stationary state solutions $\psi_i$
are obtained from the coupled system of Dirac and Klein-Gordon equations.
The variational principle leads to the Dirac equation:
\begin{equation}
\{ -i{\bf {\alpha}} \nabla + V({\bf r}) + \beta [ m* ] \}
~\psi_{i} = ~\epsilon_{i} \psi_{i}
\end{equation}
where $V({\bf r})$ represents the $vector$ potential:
\begin{equation}
V({\bf r}) = g_{\omega} \omega_{0}({\bf r}) + g_{\rho}\tau_{3} {\bf {\rho}}
_{0}({\bf r}) + e{1-\tau_{3} \over 2} {A}_{0}({\bf r})
\end{equation}
and $S({\bf r})$ is the $scalar$ potential
\begin{equation}
S({\bf r}) = g_{\sigma} \sigma({\bf r})
\end{equation}
which defines the effective mass as:
\begin{equation}
m^{\ast}({\bf r}) = m + S({\bf r})
\end{equation}
The Klein-Gordon equations for the meson fields are time-independent
inhomogeneous equations with the nucleon densities as sources.
\begin{equation}
\begin{array}{ll}
\{ -\Delta + m_{\sigma}^{2} \}\sigma({\bf r})
 =& -g_{\sigma}\rho_{s}({\bf r})
-g_{2}\sigma^{2}({\bf r})-g_{3}\sigma^{3}({\bf r})\\
\         \\
\  \{ -\Delta + m_{\omega}^{2} \} \omega_{0}({\bf r})
=& g_{\omega}\rho_{v}({\bf r}) + g_4 \omega^3({\bf r}) \\
\                            \\
\  \{ -\Delta + m_{\rho}^{2} \}\rho_{0}({\bf r})
=& g_{\rho} \rho_{3}({\bf r})\\
\                           \\
\  -\Delta A_{0}({\bf r}) = e\rho_{c}({\bf r})
\end{array}
\end{equation}

For the case of an even-even nucleus with time-reversal symmetry, the
spatial components of the vector fields, 
\mbox{\boldmath $\omega$}, \mbox{\boldmath $\rho_3$} and
\mbox{\boldmath A} vanish. For the mean-field, the nucleon spinors
provide the corresponding source terms:
\begin{equation}
\begin{array}{ll}
\rho_{s} =& \sum\limits_{i=1}^{A} \bar\psi_{i}~\psi_{i}\\
\             \\
\rho_{v} =& \sum\limits_{i=1}^{A} \psi^{+}_{i}~\psi_{i}\\
\             \\
\rho_{3} =& \sum\limits_{p=1}^{Z}\psi^{+}_{p}~\psi_{p}~-~
\sum\limits_{n=1}^{N} \psi^{+}_{n}~\psi_{n}\\
\                    \\
\ \rho_{c} =& \sum\limits_{p=1}^{Z} \psi^{+}_{p}~\psi_{p}
\end{array}
\end{equation}
where the sums are taken over the valence nucleons only. Consequently, 
the ground-state of a nucleus is obtained by solving the coupled system 
of the Dirac and Klein-Gordon equations self-consistently.
The solution of the Dirac equation is achieved by using the method of
oscillator expansion \cite{GRT.90}. In the RMF approach, the pairing is 
included within the BCS scheme, where the pairing gaps are calculated 
from the experimental masses of neigbouring nuclei.

\section{The Relativistic Hartree-Bogoliubov Theory}

Nuclei which are known to show strong pairing correlations
can be treated appropriately within the framework of the
RHB approach. This is especially important when the pairing 
correlations in the middle of a shell become important. 
However, the pairing correlations for nuclei 2 neutrons 
less or more than a magic number can be expected to be similar
in the RHB and the BCS approach. This is due to the reason that 
the pairing correlations in such nuclei are reduced to a minimal level. 

The RHB approach becomes more important while dealing with nuclei 
far away from the stability line and in particular for nuclei in the 
vicinity of the drip lines, whereby the Fermi level is usually very
near the continuum. The coupling between the bound states and the
states in the continuum is taken into account in the RHB theory. 
Thus, the RHB is a preferred scheme over the BCS in such cases. 
This seems to be necessary because we consider Ni nuclei also about and 
above N=50, which are already far away from the stability line. Moreover, 
the Sn isotopes being open-shell nuclei are known to exhibit strong 
pairing correlations. Thus, we employ the RHB theory as being the most 
suitable one for the purpose.

Analogous to the Hartree-Fock-Bogoliubov (HFB) theory \cite{Doba.84},
the RHB theory contains the self-consistent field $\hat\Gamma$ which 
represents the particle-hole correlations, and a pairing field 
$\hat\Delta$ which includes all particle-particle correlations. 
For the self-consistent mean-field, the RHB equations are given by:
\begin{eqnarray}
\label{equ.2.2}
\left( \matrix{ \hat h_D -m- \lambda & \hat\Delta \cr
                -\hat\Delta^* & -\hat h_D + m +\lambda} \right) 
\left( \matrix{ U_k({\bf r}) \cr V_k({\bf r}) } \right) =
E_k\left( \matrix{ U_k({\bf r}) \cr V_k({\bf r}) } \right).
\end{eqnarray}
where $\hat h_D$ is the single-nucleon Dirac Hamiltonian and $m$ is 
the nucleon mass. 
$U_k$ and $V_k$ are the two Dirac spinors and $\hat \Delta$
represents the pairing field. These equations are solved 
self-consistently, whereby the potentials are obtained in the 
mean-field approximation from the Klein-Gordon equations (8).

The corresponding source terms in the Klein-Gordon equations are 
sums of bilinear products of baryon amplitudes
\begin{eqnarray}
\label{equ.2.3.e}
\rho_s({\bf r})&=&\sum\limits_{E_k > 0} 
V_k^{\dagger}({\bf r})\gamma^0 V_k({\bf r}), \\
\label{equ.2.3.f}
\rho_v({\bf r})&=&\sum\limits_{E_k > 0} 
V_k^{\dagger}({\bf r}) V_k({\bf r}), \\
\label{equ.2.3.g}
\rho_3({\bf r})&=&\sum\limits_{E_k > 0} 
V_k^{\dagger}({\bf r})\tau_3 V_k({\bf r}), \\
\label{equ.2.3.h}
\rho_{\rm em}({\bf r})&=&\sum\limits_{E_k > 0} 
V_k^{\dagger}({\bf r}) {{1-\tau_3}\over 2} V_k({\bf r}),
\end{eqnarray}
where the sums are taken over all positive energy states. For M
degrees of freedom, one chooses the M positive eigenvalues $E_k$ 
for the solution that corresponds to a ground state of a nucleus 
with an even particle number. The negative energy solutions are 
ignored as it is forbidden to occupy the levels $E_k$ and
$-E_k$ simultaneously. 

The integral operator $\hat\Delta$ in eq. (10) acts on the 
wave function $V_k({\bf r})$:
\begin{equation}
\label{equ.2.4}
(\hat\Delta V_k)({\bf r}) 
= \sum_b \int d^3r' \Delta_{ab} ({\bf r},{\bf r}') V_{bk}({\bf r}'). 
\end{equation}
For the pairing field, the kernel of the integral operator is given by
\begin{equation}
\label{equ.2.5}
\Delta_{ab} ({\bf r}, {\bf r}') = {1\over 2}\sum\limits_{c,d}
V_{abcd}({\bf r},{\bf r}') {\bf\kappa}_{cd}({\bf r},{\bf r}').
\end{equation}
where $a,b,c,d$ denote all quantum numbers, except the
coordinate $\bf r$, that specify the single-nucleon states.
Here $V_{abcd}({\bf r},{\bf r}')$ denotes  matrix elements of a
general two-body pairing interaction, and 

\begin{equation}
{\bf\kappa}_{cd}({\bf r},{\bf r}') = 
\sum_{E_k>0} U_{ck}^*({\bf r})V_{dk}({\bf r}').
\end{equation}
is the pairing tensor. 

The system of the Dirac-Hartree-Bogoliubov equations (10-17) is 
solved self-consistently. 
The eigensolutions of the RHB equations (10) form a set of
orthonormal single quasi-particle states and the corresponding
eigenvalues are the single quasi-particle energies. The basis of
the quasi-particles is then transformed into the canonical basis of
the single-particle states. This basis determines the single-particle 
energies and occupations probabilities corresponding to the ground-state 
of a nucleus.

It was realized earlier that using the standard parameter sets of
the RMF theory for the pairing, one obtains highly unrealistic pairing
correlations \cite{Kucha.91}. Due to this reason, a completely 
microscopic derivation of a pairing interaction from the first 
principles has not been feasible so far. Instead, the finite-range 
pairing force of the Gogny-type is used in the p-p channel as 
suggested in ref. \cite{Gon.96}. The pairing interaction is approximated 
by a two-body finite range force of the Gogny type,
\begin{equation}
V^{pp}(1,2)~=~\sum_{i=1,2}
e^{-(( {\bf r}_1- {\bf r}_2)
/ {\mu_i} )^2}\,
(W_i~+~B_i P^\sigma 
-H_i P^\tau -
M_i P^\sigma P^\tau),
\end{equation}
with parameters $\mu_i$, $W_i$, $B_i$, $H_i$ and $M_i$
$(i=1,2)$. We use the parameter set D1S \cite{Berger.84}
for the pairing force. The Gogny force is a sum of two Gaussians
with finite range. It has been shown \cite{Berger.84} that the Gogny
force is able to represent the pairing properties of a large
number of finite nuclei including the Sn and Pb isotopes.

\section{\bf Details of the Calculations}

We have performed the RMF+BCS and RHB calculations for Ni and 
Sn nuclei with a spherical symmetry. For both the RMF+BCS and RHB 
calculations, the wavefunctions are expanded into the oscillator 
basis. The number of shells taken into account is 20 for the fermionic 
as well as bosonic wavefunctions. 

In this study, we have used the forces NL-SH and NL3 for the 
Lagrangian with the non-linear scalar self-coupling. In a large number
of studies it has been shown that the force NL-SH \cite{SNR.93} is 
able to provide a very good description of the ground-state 
properties of nuclei all over the periodic table. 
It has been found to be especially useful for exotic nuclei near 
drip lines. Recently, the force NL3 \cite{Lala.97} has been proposed 
with a view to modify the compression modulus of nuclear matter. Studies 
have shown \cite{Lalaz.98} that the force NL3 describes the ground-state 
properties of nuclei as well as NL-SH and that NL3 results are found 
to be very similar to the NL-SH ones.

We have also used the forces NL-SV1 and NL-SV2 \cite{SH.99} for the 
Lagrangian with the scalar and vector self-coupling. 
The most important point in the development of the forces NL-SV1 and
NL-SV2 is to soften the equation of state (EOS) for the nuclear
matter as compared to that with the scalar self-coupling only and to 
improve the description of the ground-state properties of nuclei
vis-a-vis the forces with the scalar self-coupling. We have also used 
the force TM1 \cite{Suga.94} for a comparative view of the results.

\section{\bf Results and Discussion}

We have chosen the chain of Ni and Sn isotopes for the present study. 
The Ni isotopes are known to be spherical in the neighbourhood of the 
stability line and this trend continues well up to the neutron shell 
closure N=50.  The Ni isotopes are favourable for the study of shell 
effects, as the shells in these nuclei are spread broadly. The large 
shells gaps are conducive to exploring the shell effects. The effects 
due to the large shell gaps appear more prominently in the 
experimental observables such as 2-neutron separation energies. 
This is exemplified by a large value of about 32 MeV of the 2-neutron 
separation energy, S$_{2n}$,  for the nucleus $^{56}$Ni. With the 
addition of a pair of neutrons to form $^{58}$Ni, the S$_{2n}$ value 
shows a dramatic decrease to $\sim$ 22 MeV in the experimental value. 
This large difference in S$_{2n}$ values of the Ni isotopes should
serve as a good test-bench to examine the shell effects. 

For heavier mass nuclei such as Sn isotopes, the shells become compressed
and consequently the $S_{2n}$ values show a lesser sensitivity to
the shell effects. In the present case, we examine the shell
effects in Sn nuclei at N=82. The experimental binding energies of 
of the neutron-rich nuclei $^{132}$Sn, $^{134}$Sn and $^{136}$Sn are 
available. This makes it possible to look into the shell effects 
at N=82 in Sn nuclei away from the stability line. 

\subsection{Ni Isotopes}

\subsubsection{Binding Energies}

We show the binding energies of the Ni isotopes with the
forces in the non-linear scalar coupling of the $\sigma$-meson.
The binding energies ($B_{th}$) of the Ni isotopes as compared to the 
experimental values ($B_{exp}$) \cite{Audi.95} are shown in Fig. 1.
The results of the RMF theory (Fig. 1a) with the BCS pairing are 
compared to those with Relativistic Hartree-Bogoliubov (RHB) 
approach using the self-consistent finite-range pairing (Fig. 1b). 
For the BCS calculations, the proton number Z=28 is assumed to be a 
closed shell and hence the proton pairing gap is taken to be zero.
The neutron pairing gap is calculated using the experimental binding 
energies of neighbouring nuclei. However, for the magic nuclei with 
N=28 and N=50, the neutron pairing gap is taken to be zero. 

The results in Fig. 1a show that the RMF+BCS approach reproduces
the binding energies of the most of the Ni isotopes well. Only for a few
nuclei about A=62 (N=34), the RMF theory underestimates the binding
energy by about 3-4 MeV ($\sim 0.5\%$). A comparison of the results of 
NL-SH and NL3 shows that the binding energies with NL3 are similar to 
those of NL-SH and that both the forces show a similar pattern of 
behaviour. The binding energies with NL-SH are slightly better than 
those with NL3, especially in the region A=50-60. 

The RHB results with the forces NL-SH and NL3 are shown
in Fig. 1b. It can be seen that the RHB approach with the self-consistent
pairing produces results which are very similar to those with the
RMF+BCS ones. The whole pattern of the BCS+RMF results is reproduced
by the RHB, with both the forces NL-SH and NL3. Moreover, the RHB results
show an improved agreement for nuclei in the region A=62. A comparison
of the results with NL-SH and NL3 shows that both the results are
very similar to each other. For the region A=70-78, both the forces 
overestimate the experimental binding energies slightly in the RHB
approach. 

With the non-linear vector self-coupling of the $\omega$-meson, 
we have used the forces NL-SV1 and NL-SV2 \cite{SH.99}. 
We have also included the force TM1 \cite{Suga.94}.
The binding energy difference ($B_{th} -B_{exp}$) is shown for the
RMF+BCS (Fig. 2a) and the RHB approach (Fig. 2b) for the three forces. 
For the BCS pairing we have used the experimental pairing gaps obtained
from the masses of neighbouring nuclei as mentioned above. The results
(Fig. 2a) show that the behaviour of the binding energies with the 
forces with the scalar-vector self-coupling is similar to those with the 
scalar self-coupling only, except for a few minor differences. The region
about A=60 shows some deviations from the experimental values.
However, the degree of agreement with the experimental data varies
with the forces. The force TM1 underestimates the binding energies by
about 4 MeV for several nuclei in the region A=54-62. The force NL-SV1
shows an improvement in the predictions of energies of these nuclei
over those of TM1. In comparison, the force NL-SV2 provides the best 
description of the binding energies, where the deviations from 
the experimental energies are reduced considerably. For the heavier Ni 
isotopes with A=68-78, the results of NL-SV1 and NL-SV2 are very 
similar, whereas for TM1 deviations from the experimental energies 
start increasing at A=74. 

We have incorporated the vector self-coupling of the $\omega$-meson
in the RHB approach for the first time. The RHB results (Fig. 2b) with 
the forces with the scalar-vector self-coupling show a high degree of 
similarity with the corresponding BCS results, as seen also for 
the non-linear scalar coupling. Only a slight improvement is noticed 
in the RHB predictions. With the force NL-SV2 the deviations from the 
experimental binding energies are improved slightly in the RHB over the
BCS for the region A=52-64. For the heavier isotopes, RHB shows
marginally higher deviations than in the BCS with NL-SV2. The trend of
the binding energies as a function of the mass number is smoothened 
in the RHB.

The RMF+BCS and RHB results of Figs. 1 and 2 demonstrate that the BCS
results are able to mock the RHB results provided experimental pairing
gaps are used. Secondly, the scalar-vector self-coupling force NL-SV2
provides a description of the binding energies which is considerably
improved over that with the non-linear scalar self-coupling only. 

\subsubsection{Two-Neutron Separation Energies}

The 2-neutron separation energies have been tabulated and plotted  
\cite{Bor.93} over the whole periodic table using the experimental
masses of nuclei. The salient feature in the curves of $S_{2n}$ values
in ref. \cite{Bor.93} is a kink which can be observed conspicuously 
about all the major magic numbers. A kink implies that $S_{2n}$ 
values for nuclei with 2 neutron numbers above a magic number 
exhibit a dramatic decrease as compared to a nucleus with the magic 
number or nuclei below the magic number. This sudden decrease in 
the $S_{2n}$ values signifies a shell gap above the magic number 
and thus reflects the role played by the ensuing shell effects. 
The magnitude of a shell gap does also show a significant dependence 
on mass number of a nucleus. It is thus expected that the shell effects
vary strongly depending upon the part of the periodic table one is
considering.

Figure 3 shows the results obtained with the forces with the non-linear 
scalar self-coupling NL-SH (Fig. 3a) and NL3 (Fig. 3b). The results for 
the RMF+BCS and the RHB are compared with the experimental $S_{2n}$ 
values (solid circles). A staggeringly high value of $\sim$ 35.0 MeV 
for the 2-neutron separation energy for the nucleus $^{52}$Ni can be seen. 
It decreases smoothly to 30.8 MeV for the doubly magic 
nucleus $^{56}$Ni. However, on adding 2-neutrons to the magic core, 
the $S_{2n}$ value shows a dramatic decrease to 22.5 MeV for $^{58}$Ni. 
This difference in the $S_{2n}$ values is a clear measure of the shell
gap at the magic number. Inevitably, the downward kink at N=28
provides a sensitive observable for an investigation of the shell 
effects in nuclei. Therefore, we focus upon the 2-neutron separation 
energies to analyze the  characteristics of the shell effects. 

A comparison of the BCS results with the RHB in Fig. 3 shows that 
except for a few minor differences, the two approaches provide results 
which are very close to each other for most of the isotopic chain. 
This is also the case in the extremely neutron-rich region. Essentially, 
these results mirror the accord in the binding energies in the two 
approaches as shown in Fig. 1. This implies that the experimental 
pairing gaps used in the BCS approach are adequately represented 
by the self-consistent Gogny pairing in the RHB approach.

For a comparative look at the various predictions, $S_{2n}$ values 
for RMF+BCS are shown in Fig. 4a and for RHB in Fig. 4b. A striking 
similarity is observed in the results of Fig. 4a and Fig. 4b. It arises 
naturally due to the similarity in the predictions of the binding 
energies with RMF+BCS and RHB as already observed in Fig. 1. There are, 
however, two differences in the RHB results as compared to the BCS ones.
Firstly, the inclusion of the self-consistent pairing in RHB increases 
the kink at N=28 slightly. Secondly, the RHB results show an improved 
agreement with the experimental data in the region A=64-70 as compared to
the BCS results. 

The RHB results employ the self-consistent pairing and ought to be 
considered as more realistic than BCS. Therefore, we focus our 
attention primarily upon the RHB results. As observed for the binding 
energies in Fig. 1, the $S_{2n}$ values with RHB for NL-SH and NL3 
are also close to each other. As far as the experimental data is 
concerned, the RHB is able to describe the $S_{2n}$ data on the 
lighter (A=52-56) Ni isotopes very well. However, for nuclei just 
above the magic number N=28, i.e., for $^{58}$Ni and $^{60}$Ni, 
the RHB values are lower than the experimental values by about 
2-2.5 MeV. This implies that the shell gap at N=28 is higher in the 
theory than indicated by the experimental data. Thus, the RMF theory 
with the non-linear scalar coupling shows shell effects which are 
stronger than suggested by the experimental data. 

It can be seen from Fig. 4b that for nuclei heavier than $^{62}$Ni, 
the $S_{2n}$ values for NL-SH are closer to the experimental data 
than the NL3 values. The $S_{2n}$ values for nuclei with A=70-78 are 
reproduced successfully both by NL-SH and NL3. It may be noted that
the empirical data on $S_{2n}$ exist only up to $^{78}$Ni. The mass of 
the nucleus $^{80}$Ni is not known experimentally. Therefore, our 
knowledge about the shell effects far away from the stability line 
is severely constrained. In this context, the nucleus $^{80}$Ni 
constitutes one such nucleus which has a potential to expose shell
effects in reaching out to the neutron drip-line. A measurement of the
mass of the nucleus $^{80}$Ni would undoubtedly shed light on how the 
shell effects behave far away from the stability line. Here, we
provide our predictions only on this account. 

Due to uncertainties of pairing and coupling of 
states to the continuum, the BCS results are not expected to be fully
realistic in this region. The RHB approach is the most appropriate one 
for the highly exotic Ni nuclei. The presence of a kink in the 
RHB results (Fig. 4b) at N=50 signifies the existence of shell effects 
which are still strong far away from the stability line. Comparatively, 
the shell effects with NL-SH are slightly stronger than with NL3 in the 
extreme region. We will also compare the shell effects in the RHB approach 
to those in the Hartree-Fock Bogoliubov (HFB) approach with the Skyrme 
force SkP in the latter part of this paper. 
 
The results on 2-neutron separation energy for the model with the 
scalar-vector self-coupling are presented in Fig. 5. Here we show the 
results with the forces NL-SV2 and TM1. The upper panel (Fig. 5a) displays 
the $S_{2n}$ values with the BCS pairing. A comparison of the BCS results 
with the RHB results (Fig. 5b) in the region of experimentally known nuclei
shows that the two approaches give results which are remarkably similar, 
provided the experimental pairing gaps are used in the BCS calculations. 
 
Again, we focus upon the RHB results due to the self-consistency of the 
pairing. The results with NL-SV2 show a good agreement with
the experimental data for most of the isotopic chain. 
For the region below N=28, the $S_{2n}$ values with NL-SV2 are closer 
to the experimental data than those with TM1. The relative difference in 
the $S_{2n}$ values of nuclei $^{56}$Ni and $^{58}$Ni with TM1 is less than 
the experimental data, whereas it is closer to the experimental difference 
for the force NL-SV2. In the region A=62-68, where deviations from the 
experimental data are most notable, the NL-SV2 results are closer to the 
experimental data than TM1. This is due to a better degree of agreement 
with the experimental binding energies obtained by using the vector 
self-coupling force NL-SV2 as shown in Fig. 2.

\subsubsection{Shell Effects}

We compare in Fig. 6 the $S_{2n}$ values obtained with 
the scalar-self coupling (NL-SH) to those with the 
scalar-vector self-coupling (NL-SV2). In Fig. 6a
the $S_{2n}$ values across the shell-closure N=28 are presented.
A comparison with the experimental data about N=28 shows that the force 
NL-SH gives shell effects which are stronger than suggested by the empirical
data. The relative difference between the $S_{2n}$ values of $^{56}$Ni and
$^{58}$Ni is 11.1 MeV with NL-SH and 9.2 MeV with NL-SV2. 
The difference in the $S_{2n}$ values of nuclei about a shell closure
is an indicator of the shell gap at the magic number. The results show 
that the shell gap at N=28 is reduced considerably with the scalar-vector 
self-coupling (NL-SV2) as compared to the scalar coupling. In comparison, 
the experimental difference amounts to about 8.4 MeV. Thus, the shell gap 
at N=28 with NL-SV2 is very close to the experimental one.

Shell effects with the Hartree-Fock Bogoliubov approach using the
Skyrme force SkP have been discussed often in the literature 
\cite{chen.95,pear.96,kratz.98}. In order to clarify the situation,
we include the $S_{2n}$ values \cite{doba_pri.94} obtained in the HFB 
approach with the force SkP in Fig. 6a. It is seen clearly that the 
SkP values cross the shell-closure N=28 (A=56) smoothly. This has an 
unambiguous implication that {\it contrary} to the experimental 
observation, the shell effects with SkP are quenched strongly at N=28
and consequently the shell gap at N=28 is diminished considerably. 
This aspect of SkP will become obvious in the following where we 
compare single-particle levels from various approaches.

The 2-neutron separation energies across the N=50 shell closure are shown
in Fig. 6b. The experimental data on $S_{2n}$ is available only up to 
A=78 (N=50) as shown by the solid circles. 
How the shell effects behave about N=50 in the
region very far away from the stability line is still an open question ?
We will show elsewhere \cite{Farhan.99} that the very few experimental data 
available for N=50 are consistent with strong shell effects. 
Determination of the nature of shell effects for waiting-point nuclei in 
the vicinity of N=50 are crucial to understanding the r-process 
nucleosynthesis in the neighbourhood of A=80. A future measurement of mass 
of the nucleus $^{80}$Ni would help resolve this issue. In the present work, 
we make our predictions on the basis of the extrapolation from the region of 
stability. A comparison of the $S_{2n}$ values for $^{78}$Ni (N=50) and 
$^{80}$Ni for the forces NL-SH and NL-SV2 show that the shell gap at 
N=50 is larger for NL-SH than for NL-SV2. This situation is very 
similar to that we have seen in the known region of N=28. Thus, also at 
N=50 the shell effects with NL-SH are stronger than NL-SV2. Taking 
into account a good agreement of NL-SV2 results with the experimental 
data about N=28, we predict that at N=50 the shell effects still
remain strong. This is in accordance with a general perception that
the magic number N=50 constitutes a strong shell closure. This feature
is thus predicted to persist also far away from the stability line. 
These results are contrasted sharply (Fig. 6b) with the results of the
HFB with the Skyrme force SkP which predicts a smooth decrease in the 
$S_{2n}$ values across the magic number N=50. The strong quenching of 
the shell effects at N=50 with SkP as seen here is consistent with the
similar quenching shown by this force also at N=28 (see Fig. 6a).

\subsubsection{Single-Particle Levels}

Differences in the shell effects in various approaches are exhibited 
succinctly in the single-particle spectrum as shown in Fig. 7. The neutron 
single-particle spectra for the nucleus $^{58}$Ni are displayed 
for the forces NL-SH and NL-SV2 and a comparison is made with SkP. 
The force NL-SV2 with the scalar-vector self-coupling shows a decrease 
in the shell gap at N=28 as compared to NL-SH indicating again that 
the shell effects with NL-SV2 are softer. This is is conformity with the 
corresponding $S_{2n}$ values in Fig. 6a. 

The single-particle spectrum with SkP in Fig. 7 reveals a striking
difference with that of NL-SH and NL-SV2. The levels with SkP are
compressed significantly as compared to NL-SH and NL-SV2. 
The mean-field potential with SkP appears to be much shallower than the 
corresponding one in the RMF theory. The most important difference about 
the shell effects is that with SkP the shell gap at N=28 is reduced 
significantly. This is the reason that a strong quenching of the shell
effects is exhibited by SkP at N=28.

With a view to illustrate the shell effects at N=50, the single-particle 
levels for the nucleus $^{80}$Ni are shown in Fig. 8 for NL-SH, NL-SV2
and SkP. The shell gap at N=50 with NL-SV2 is reduced  as compared to 
that with NL-SH. This reduction in shell gap is similar to that also 
observed for N=28 with NL-SV2 in Fig. 7. Thus, the behaviour of the 
single-particle levels for $^{58}$Ni and $^{80}$Ni and the reduced 
shell gaps at N=28 and N=50 serve to illustrate the point that 
the slightly-softer shell effects with NL-SV2 are consistent with 
the experimental data. 

The single-particle spectrum for $^{80}$Ni with SkP in Fig. 8 is strikingly 
different than that for NL-SH and NL-SV2. The levels with SkP are compressed
significantly as compared to NL-SH and NL-SV2. This feature is very similar
to that portrayed in Fig. 7. In fact, such a feature seems to be a generic one
for SkP. The single-particle spectrum for SkP shows that the shell gap at 
N=50 with SkP is about a factor of 2 smaller than the corresponding one with
NL-SV2. Therefore, with SkP the shell closure N=50 is quenched strongly.
This is consistent with the smooth decrease of $S_{2n}$ values with
SkP across N=50 (see Fig. 6b). 

The strikingly different single-particle scheme with SkP may be attributed 
to the effective mass $m^* = 1$ assumed for this force. The effective mass
has an important influence on the single-particle structure of nuclei.
The quantities which lend their influence to a single-particle scheme are
the effective mass and the compressibility of nuclear matter. The latter
is intricately connected to the saturation density through the properties
of finite nuclei. Recently, it has been shown that in the density
dependent Skyrme approach, the shell effects show a significant
dependence on the compressibility of nuclear matter \cite{Riken.99}. 

In  Fig. 9 we examine predictions on the shell effects from  various mass 
models. The $S_{2n}$ values are shown for the mass formulae the Finite-Range 
Droplet Model (FRDM) \cite{FRDM.95} and the Extended Thomas-Fermi with 
Strutinsky Integral (ETF-SI) \cite{Abou.95}. The FRDM data about N=28 
(Fig. 9a) are in close agreement with the experimental data, showing
that the shell effects about N=28 are represented adequately in the FRDM. 
On the other hand, in the ETF-SI which is based upon the Skyrme Ansatz, 
the shell effects are weakened significantly at N=28. A comparison of the 
ETF-SI data with the SkP predictions shows that in both the cases, 
the shell effects are quenched strongly. 
It may be recalled that in the ETF-SI approach, shell corrections to a 
smooth component of energy-functional are added using the Strutinsky 
method. Thus, the nature of the shell corrections as obtained in this 
approach is crucial to deciding the strength of the shell effects.

At N=50 (Fig. 9b) predictions of the ETF-SI for the shell effects are, 
however, in stark contrast to those of SkP. Whilst SkP predicts a
washing out of the shell effects at N=50 as discussed above, the ETF-SI, 
on the other hand, shows shell effects which are strong at N=50. 
This difference between ETF-SI and SkP at N=50 is not easy to understand 
notwithstanding the similar quenching at N=28 in both the  approaches. 
The strength of the shell effects with SkP derives from the high value of
the effective mass. This tends to reduce shell gaps at most of the magic
numbers. On the other hand, shell effects in ETF-SI stem from as to how 
the shell effects are evaluated in different regions. Shell corrections
in the ETF-SI based upon the experimental data in the region maintain
a strong shell closure at N=50. A point of comparison, however, may be
made here that the magnitude of the shell gap as reflected by the 
$S_{2n}$ values at N=50 in ETF-SI is similar to that projected by the 
force NL-SV2. The FRDM results, however, show shell effects which are 
stronger than those predicted by both ETF-SI and NL-SV2. 

\subsection{Sn Isotopes}

The Sn isotopes provide a very long chain where the experimental masses are
known over a large range of isospin. The experimental data is presently 
available above the neutron magic number N=82 up to the mass A=136 
(N=86). This allows us to examine the shell effects about the neutron 
shell closure N=82 which in the present case lies reasonably away 
from the stability line. The nature of the shell effects about N=82 
should reveal as to how the shell closure N=82 retains (or looses) 
it magic character in the region far away from the stability line.

We have included Sn isotopes with mass A=100-138 for the present
study. It must be pointed out that the Sn isotopes have long been
known to possess strong pairing properties akin to superfluidity.
Thus, the role of the pairing is very important for the Sn nuclei.
As in case of the Ni isotopes, we have performed RMF+BCS calculations using
the pairing gap obtained from the experimental masses of neighbouring
nuclei. We have also performed the RHB calculations using the
self-consistent finite-range pairing. 

\subsubsection{Binding Energies}

The binding energy differences $(B_{th} - B_{expt})$ obtained with the
two approaches are shown in Fig. 10. A direct comparison is made
between the results of the forces NL-SH and NL3. The RMF+BCS results
(Fig. 10a) show  strong deviations from the experimental binding energies
in the vicinity of A=100. For the other regions of the range,
the NL-SH results are generally better than NL3. However, in the
vicinity of the magic number N=82 (A=132), NL3 predictions are slightly 
better than those of NL-SH. The BCS results show several undulations
which arise from the inadequacy of the BCS pairing. A kink due to the
similar reasons is seen at A=108 with both the forces.

The RHB results with the self-consistent pairing (Fig. 10b) are free from
the fluctuations of Fig. 10a. The binding energies show a smooth trend
with mass number and the kink at A=108 has disappeared. 
The smoothness of the RHB results vis-a-vis the BCS ones is an
indication that the self-consistent pairing is very important for
nuclei like Sn isotopes. 

The RHB results (Fig. 10b) with NL-SH and NL3 follow a similar pattern in 
the binding energy. It is observed that for the light Sn 
isotopes, NL-SH overestimates the binding energies by a few MeV. 
This aspect is especially apparent at the doubly magic nucleus 
A=100 both for NL-SH and NL3. In the region A=100-110,
the NL3 values show a better agreement with the experimental data
than NL-SH. However, in the latter part of the chain where stable
Sn isotopes exist, the NL-SH and NL3 results are very similar, 
showing a good agreement with the experimental data within 
1-2 MeV. In the neighbourhood of A=132 disagreement of about 2-3 MeV 
with the experimental data appear for both NL-SH and NL3.

For a direct comparison of the RMF+BCS with the RHB,
we show in Fig. 11 the results from the two approaches for
the forces NL-SH (a) and NL3 (b). The RHB results show a smooth
behaviour with the mass number for both the forces. The change 
from BCS to RHB is moderate for NL-SH. However, for NL3 results
the change from BCS values to the RHB values is much larger
than NL-SH. Thus, the self-consistent pairing brings about
a significant change in the energies of NL3 especially in
the light-mass Sn nuclei. This shows that with the force NL3,
the experimental pairing gaps are not able to take into account
the pairing correlations correctly for the Sn nuclei in 
particular those with lower mass. On the whole, this situation of 
Sn isotopes is unlike the results of Ni isotopes (see Figs. 1 and 2) 
where the RHB results mirrored much the results from the BCS. 
In other words, within the BCS the experimental pairing gaps 
for the Ni isotopes were adequate to represent the pairing
correlations. 
 
In Fig. 12 we show the results with the forces with the scalar-vector 
self-couplings, with both the RMF+BCS and the RHB theory.
Here the forces NL-SV1, NL-SV2 and TM1 are included. The presence of 
a kink at A=108 in the BCS calculations indicates that in spite
of taking the pairing gaps from the experimental data, there seems to be
an inconsistency in representing the pairing correlations appropriately
in the BCS formalism. The smoothness of the RHB results with the
self-consistent pairing shows that this drawback is eliminated in 
the RHB approach. This is true for all the forces included here.
Examining the the BCS results only, it is seen that the binding energies 
with NL-SV1 and TM1 are closer to the experimental data than NL-SV2
for the light-mass Sn nuclei. NL-SV2 results in this region overpredict the
experimental data slightly. For the region A=110-130 the binding energies
with all the forces show a good agreement with the data within about
1 MeV. In general, the BCS results with NL-SV1 show an overall good
agreement with the data, except about the doubly magic nucleus $^{100}$Sn.
In comparison, the TM1 results show a slightly higher disagreement 
with the data. The divergence of TM1 with the data is, however, more 
pronounced in the extreme region about A=136.

We have compared the RHB results with the scalar-vector self-coupling 
forces in Fig. 13. It is seen that the results of NL-SV2 and TM1 are similar
over most of the mass range except in the vicinity of the doubly
magic nucleus $^{100}$Sn. Both the forces overestimate the
binding energies of Sn nuclei by about 2-4 MeV. This implies that the
underlying pairing correlations with the given pairing force D1S
are stronger for the forces NL-SV2 and TM1. On the other hand,
the force NL-SV1 shows the best agreement with most of
the data. The NL-SV1 results are within 1-2 MeV of the experimental 
energies except for the region about $^{100}$Sn. In the neighbourhood
of the other doubly magic nucleus $^{132}$Sn, the NL-SV1 results are
are similar to NL-SV2 and are in agreement with the data within 
about 1-2 MeV. 

\subsubsection{Two-Neutron Separation Energies}

The 2-neutron separation energies for the Sn isotopes with the RHB 
approach using the forces with the non-linear scalar self-coupling are 
shown in Fig. 14. For both the forces NL-SH and NL3, 
there is a very good agreement of the results with the experimental
data in the mass region A=110-132. Both the forces show a slight 
divergence from the experimental values as one moves below A=110 
towards A=100. 

The results in the vicinity of the doubly magic nucleus A=132 
show that the shell gap at N=82 is larger than the experimental data  
suggests. Consequently, the shell effects with NL-SH are
stronger than the empirical data. This is similar to what was also
observed for the case of Ni isotopes at N=28. In comparison, NL3 shows 
shell effects which are slightly less stronger than those of NL-SH.
However, the experimental difference in $S_{2n}$ values of
$^{132}$Sn and $^{134}$Sn is still overestimated by NL3.

The 2-neutron separation energies obtained with the scalar-vector 
self-coupling are shown in Fig. 15. The results with all the forces are
in very good agreement with the experimental data over the whole
mass range except for nuclei in the vicinity of the doubly magic
A=132. Both NL-SV1 (a) and TM1 (c) overestimate slightly the experimental
$S_{2n}$ values for nuclei lighter than $^{132}$Sn. In comparison,
the NL-SV2 results show an impressive agreement with the data almost
over the whole mass range. For nuclei heavier than $^{132}$Sn, however,
all the forces show a small divergence with the data. 

Considering the shell gap at N=82, we observe that all the forces 
with the scalar-vector self-coupling in Fig. 15 are in good agreement 
with the experimental gap. This demonstrates that these forces reproduce 
the experimental shell gaps about N=82 very well. Further,
the presence of a well developed kink in $S_{2n}$ values at
A=132 (N=82) emphasizes that the shell effects in this region which is
away from the stability line do remain strong. Indeed, this leads to
a sharp reduction in the  $S_{2n}$ value of the isotope $^{134}$Sn.
For nuclei heavier than this, the 2-neutron separation energies show
only a mild decrease with the mass number. This is supported by the only 
experimental data point available at A=136. Theoretically, this behaviour
is expected to be continued for heavier Sn isotopes as shown by another
nucleus (A=138) in Fig. 15. It may be mentioned that the neutron drip 
line in Sn nuclei is surmised to occur at very large mass number 
about A=170. It is expected that the 2-neutron separation energy would 
decrease very slowly to a vanishing value until the drip line is reached. 

\subsubsection{Single-Particle Levels}

Figure 16 shows the neutron single-particle levels in the canonical basis
for the nucleus $^{102}$Sn as obtained with the forces NL-SH and NL-SV2.
It can be seen that with NL-SV2 the shell gap at N=50 is reduced as 
compared to NL-SH. This reduction in the shell gap near the Fermi surface
with the scalar-vector self-couplings was also observed for the Ni nuclei.
A comparison of the single-particle levels obtained with the Skyrme force SkP
shows that the level density with this force is generally higher than
the RMF forces and that due to this reason the shell gaps are
suppressed rather strongly. The shell gap at N=50 is about half 
the shell gap in the RMF theory. This is illustrative of the quenching 
with the force SkP, which is much discussed in the literature 
\cite{kratz.93,chen.95}.

We show in Fig. 17 the single-particle levels for the nucleus $^{134}$Sn.
We have chosen this nucleus in order to examine the shell gaps about
the magic number N=82. The shell gap at N=82 with NL-SV2 is smaller  
than with NL-SH and is consistent with the reduction in shell gap as 
reflected by the $S_{2n}$ values about N=82 in Fig. 15 as compared 
to NL-SH. In comparison, with the force SkP the shell gap at N=82 
is reduced significantly.

In order to review the situation on the shell effects 
at N=82 with various mass models, we present in 
Fig. 18 the $S_{2n}$ from FRDM, the ETF-SI and the mass model based 
upon infinite nuclear matter (INM) \cite{Satpathy.99}. Although the
mass formulae describe the $S_{2n}$ values of most of the Sn isotopes 
correctly, there are differences in predictions about the shell closure.
Comparison of $S_{2n}$ values in the vicinity of A=132 in Fig. 18a shows 
that the gap in the $S_{2n}$ values about N=82 is slightly smaller with
FRDM than the experimental one implying that in the FRDM the N=82 shell
closure is slightly weaker. However, the shell effects with FRDM  
are still stronger as compared to the scenerio of the strong quenching 
with SkP in Fig. 9. The ETF-SI, on the other hand, is able 
to come closer to the experimental data. Thus, at N=82 the ETF-SI 
demonstrates the shell effects which are stronger than FRDM. At N=50 the 
shell effects with ETF-SI were also found to be strong and consistent 
with the experimental data. It may be reminded that at N=28 the shell 
effects are strongly quenched in the ETF-SI as opposed to the 
experimental data.

In Fig. 18b we compare the predictions of the INM model \cite{Satpathy.99}
and SkP with the experimental data. The $S_{2n}$ value from the INM model 
for $^{134}$Sn is much higher compared to the experimental one. 
The corresponding shell gap at N=82 is consequently too small, implying 
a quenching of the shell effects. For all other isotopes, the $S_{2n}$ 
values from the INM agree well with the experimental data. 
It may be mentioned that in the INM mass formula, determination of the 
nuclear matter properties and properties of finite nuclei is achieved on 
the basis of a model independent analysis of the available experimental 
data. However, the shell effects in the model seem to be quenched as 
seen at N=82.  

The SkP results show agreement with the experimental data except for 
nuclei in the vicinity of the magic number. With SkP the shell gap 
at N=82 is reduced as compared to the experimental data. The reduction
in the shell gap at N=82 (see Fig. 17) leads to an $S_{2n}$ value 
for $^{134}$Sn, which is higher than the experimental one. 
This is indicative of a reduced
shell strength with SkP at N=82. In contrast to the situation in Ni
isotopes, it may, however, be stated that SkP does show a semblance
of some shell strength at N=82. The quenching of the shell effects at 
N=82 with SkP is not so strong as has been witnessed for the Ni 
isotopes at N=28. In comparison, a strong quenching was presented 
by SkP also at N=50. Therefore, contrary to the available experimental
data, the quenching of the shell effects seems to be a generic 
feature of the force SkP.

\section{Summary and Conclusions}

We have studied the shell effects in nuclei about the stability line
in the RMF theory. The chains of Ni and Sn isotopes have been considered,
where experimental data over a large range of isospin are available.
First, employing the existing forces NL-SH and NL3 with the non-linear 
scalar self-coupling of $\sigma$-meson, we have investigated the 
ground-state properties of the Ni and Sn isotopic chains. It is shown that
in the RMF+BCS approach where the pairing gaps are taken from the 
experimental data, the experimental binding energies of Ni isotopes 
are described well. Calculations within the RHB approach with the 
self-consistent finite-range pairing show only a slight improvement 
over the RMF+BCS results in this isotopic chain. This shows that the 
RMF+BCS approach in conjunction with the pairing gaps obtained 
from the experimental data suffices to 
bring about a good description of the binding energies of the Ni isotopes. 
However, the same can not be said for the Sn isotopes. The pairing 
correlations for Sn nuclei are not adequately described within the BCS
scheme. It is shown that for these nuclei, the RHB approach with the 
self-consistent pairing is necessary in order to get a good
agreement with the binding energies. 

In addition to the nonlinear scalar coupling of $\sigma$ meson, 
we have introduced non-linear vector self-coupling of $\omega$ meson
in the RHB approach. The self-consistent pairing is included by 
taking the finite-range Gogny force in the p-p channel. We have
employed the recently developed forces NL-SV1 and NL-SV2 and the 
force TM1. It is shown that the RHB approach with the non-linear 
scalar-vector self-couplings provides a very good description of 
the binding energies of Ni and Sn isotopes. 

We have investigated the shell effects in nuclei about the stability
line within the two model Lagrangians, i.e. the RHB theory with the non-linear
scalar self-coupling of $\sigma$-meson only and the RHB theory
with the non-linear self-couplings of both $\sigma$ and $\omega$ 
mesons. The experimental data on 2-neutron separation energies
of Ni and Sn isotopes have been used. The sharp decrease in the $S_{2n}$ 
values about a magic number is taken as a measure of the shell gap. 
The experimental data on Ni encompass the magic number N=28, whilst
the magic number N=82 is encompassed by the Sn data.
It is shown that the forces with the non-linear scalar self-coupling 
give shell gaps which are larger than suggested by the experimental data. 
This is found to be the case for N=28 in Ni isotopes as well
as for N=82 in Sn isotopes. 

The results with the Lagrangian with the scalar-vector self-couplings 
show that the force NL-SV2 describes the experimental data on the 
2-neutron separation energies very well. The shell effects about the
magic numbers in Ni and Sn isotopes with the scalar-vector self-couplings 
are shown to be consistent with the experimental data. The single-particle
spectra for the Ni and Sn nuclei in the vicinity of magic numbers
show that the shell gaps at N=28, N=50 and N=82 are smaller in the
model with the scalar-vector self-couplings as compared to the scalar 
self-coupling only. 

A comparison of the experimental data and the results of the scalar-vector
self-coupling model with the HFB results using the Skyrme
force SkP for the Ni isotopes has shown that the shell effects with 
SkP are quenched strongly at N=28. This is contrary to the experimental data
which show that the shell effects are substantially stronger. The weak
or quenched shell effects are also shown in Ni isotopes by SkP at N=50.
The quenching of the shell effects with the force SkP is demonstrated
by the corresponding single-particle spectra for $^{58}$Ni, $^{80}$Ni 
and $^{134}$Sn, whereby the shell gaps at N=28, N=50 and 
N=82, respectively are considerably smaller than
the RHB single-particle gaps. In summary, the quenching of the shell
effects seems to be an artifact of the force SkP and is not supported by
the experimental data. The substantial reduction in the shell gaps
with SkP stems from the correspondingly high level density. This is 
due to a large effective mass being chosen for this force. 

It is expected that due to inherently different single-particle
structure in each model the shell effects at the stability 
line would have varying consequences on the shell effects at the
drip lines. It was shown earlier \cite{SLH.94} that with the non-linear
scalar self-coupling in the RMF theory shell effects near the neutron 
drip line in the vicinity of N=82 were predicted to be strong. 
This conclusion was based upon RMF theory with the BCS 
pairing using the force NL-SH. It was also shown \cite{SLH.94} that the 
pairing itself does not create or destroy the shell effects.  
The conclusions of the strong shell effects near the neutron 
drip line were contested \cite{Doba.94}. It was suggested on the basis 
of HFB calculations with SkP that the shell effects near the drip line 
quench strongly. In the light of the forgoing discussions, it is only 
expected that the shell gap at N=82 with the force SkP would diminish.
Thus, the quenching of the shell effects near the drip line
at N=82 with the force SkP is in complete accord with the quenching 
shown by it near the stability line. This quenching is, however, not
supported by the experimental data at the stability line as 
demonstrated in this work. 

As shown above, the force NL-SH with the scalar coupling gives
shell effects which are slightly stronger than the experimental data.
This fact notwithstanding, we have performed a study \cite{Farhan.99}
to investigate the role of the vector self-coupling on shell effects at the 
drip line. It is observed that with the scalar-vector self-couplings which
describes the shell effects at the stability line well, shell effects 
near the neutron drip line are slightly less stronger than with 
the scalar self-coupling only. However, we do not find indications 
of any shell quenching. Details of this work will be provided elsewhere 
\cite{Farhan.99}. 

It may be noted that in the fits of the r-process abundances 
\cite{chen.95}, HFB results based upon the force SKP were used. 
It was shown \cite {chen.95} that the HFB results give an improved 
fit to the global r-process abundances. Since the force SkP is known 
to quench shell effects as shown above, it was concluded on this basis 
that the shell effects at N=82 at the neutron drip line are quenched. 
However, as discussed in our work, this quenching with SkP is not
consistent with the experimental data. Thus, an improved fit to
r-process abundances using the SkP results may be insufficient to
assert the shell quenching at N=82 near the neutron drip-line,
notwithstanding an overwhelming shell quenching with SkP.

\section{Acknowledgment}

This research was supported by the Kuwait University Research Administration
through the Project No. SP056. We thank Prof. J. Dobaczewski for supplying 
the HFB results on Ni isotopes.

\newpage
\leftline{\Large {\bf Figure Captions}}
\vskip 1 true cm
\begin{description}

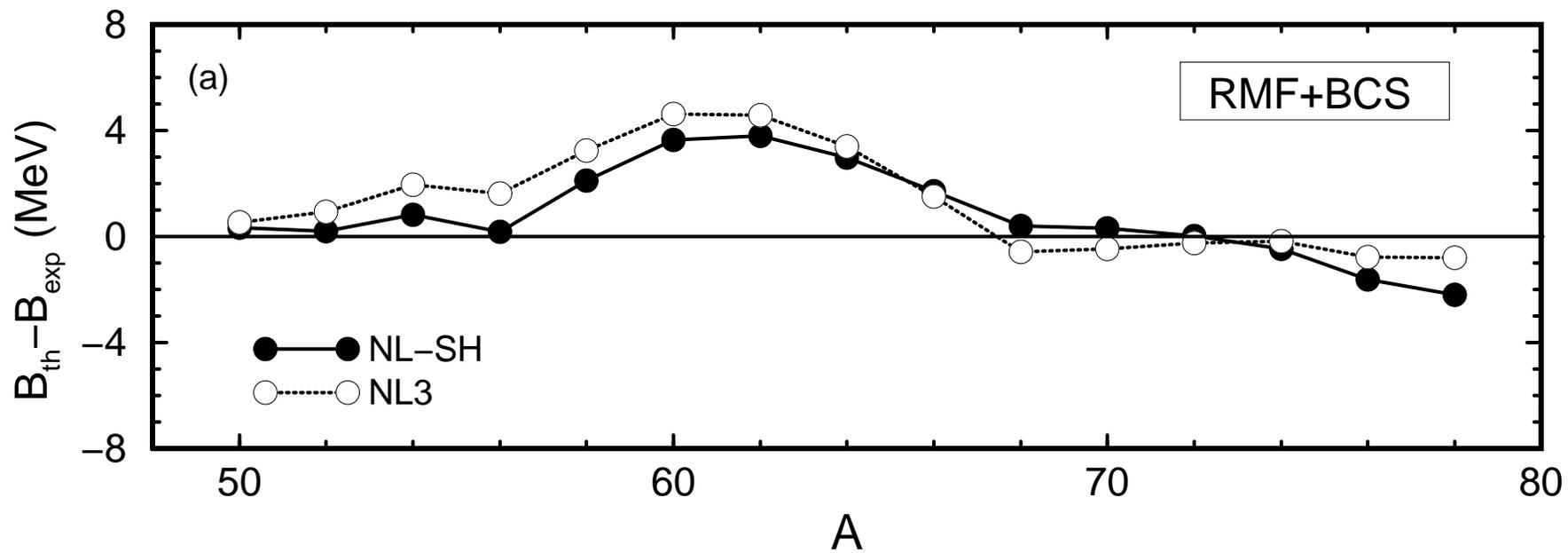
\item[Fig. 1] Binding energy of Ni isotopes with respect to the experimental
values, in (a) RMF approach with BCS pairing and (b) RHB approach with
self-consistent finite-range pairing, using the forces NL-SH and NL3 with 
the non-linear scalar self-coupling of $\sigma$ meson.

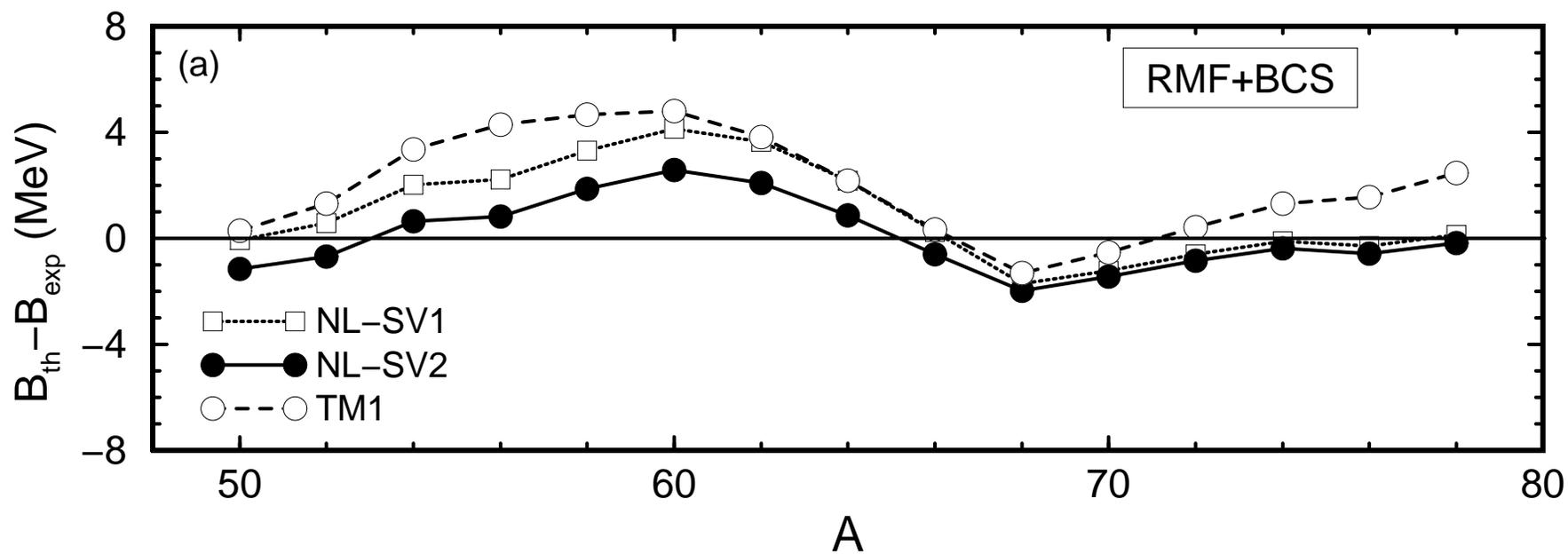
\item[Fig. 2] The same as in Fig. 1, with the forces NL-SV1, NL-SV2 and
TM1 with the non-linear scalar-vector self-coupling.

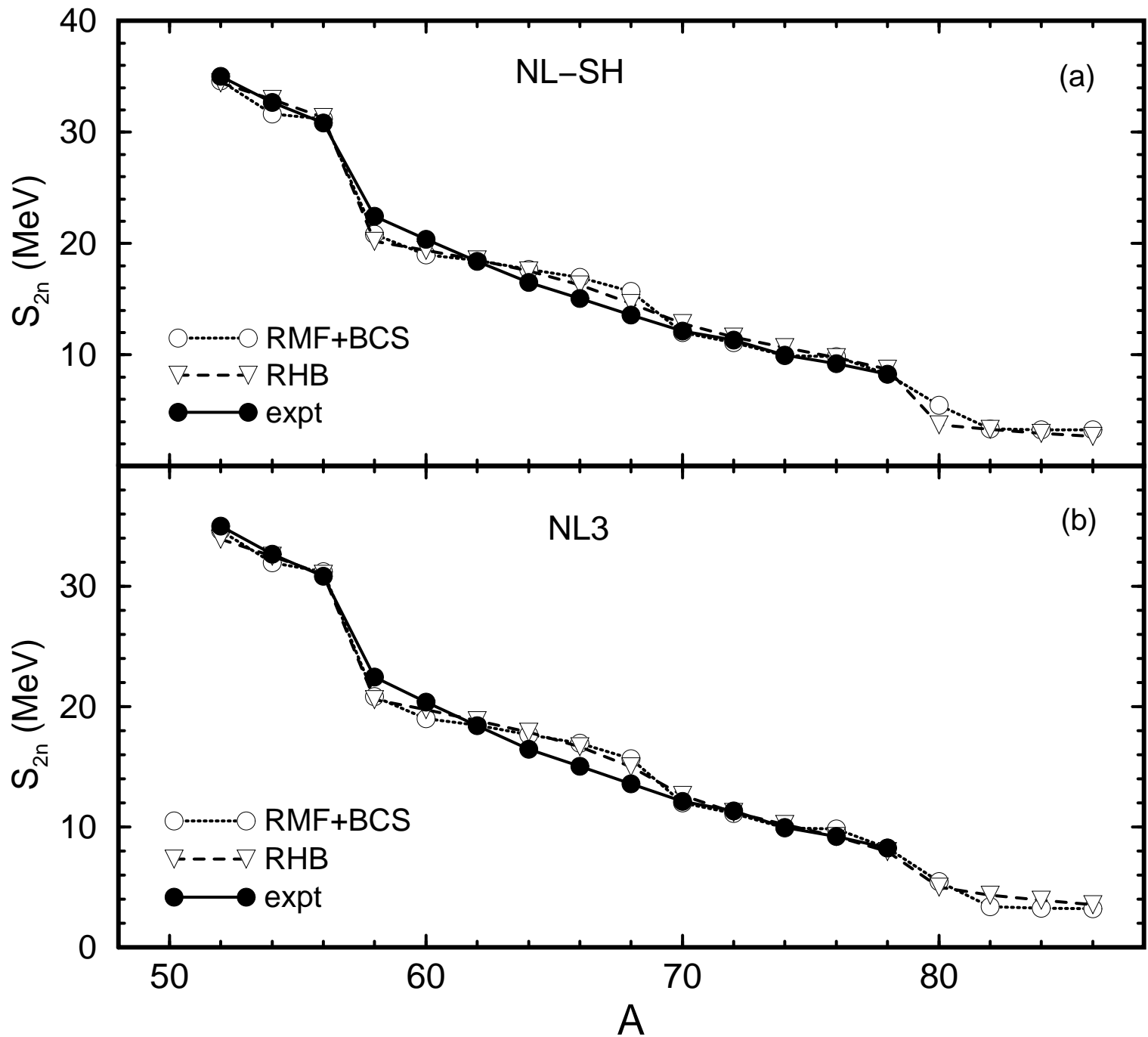
\item[Fig. 3] The 2-neutron separation energy $S_{2n}$ of Ni isotopes
using the forces with the non-linear scalar self-coupling (a) NL-SH and 
(b) NL3. A comparison of the RMF+BCS and RHB values is made with the 
experimental data. 

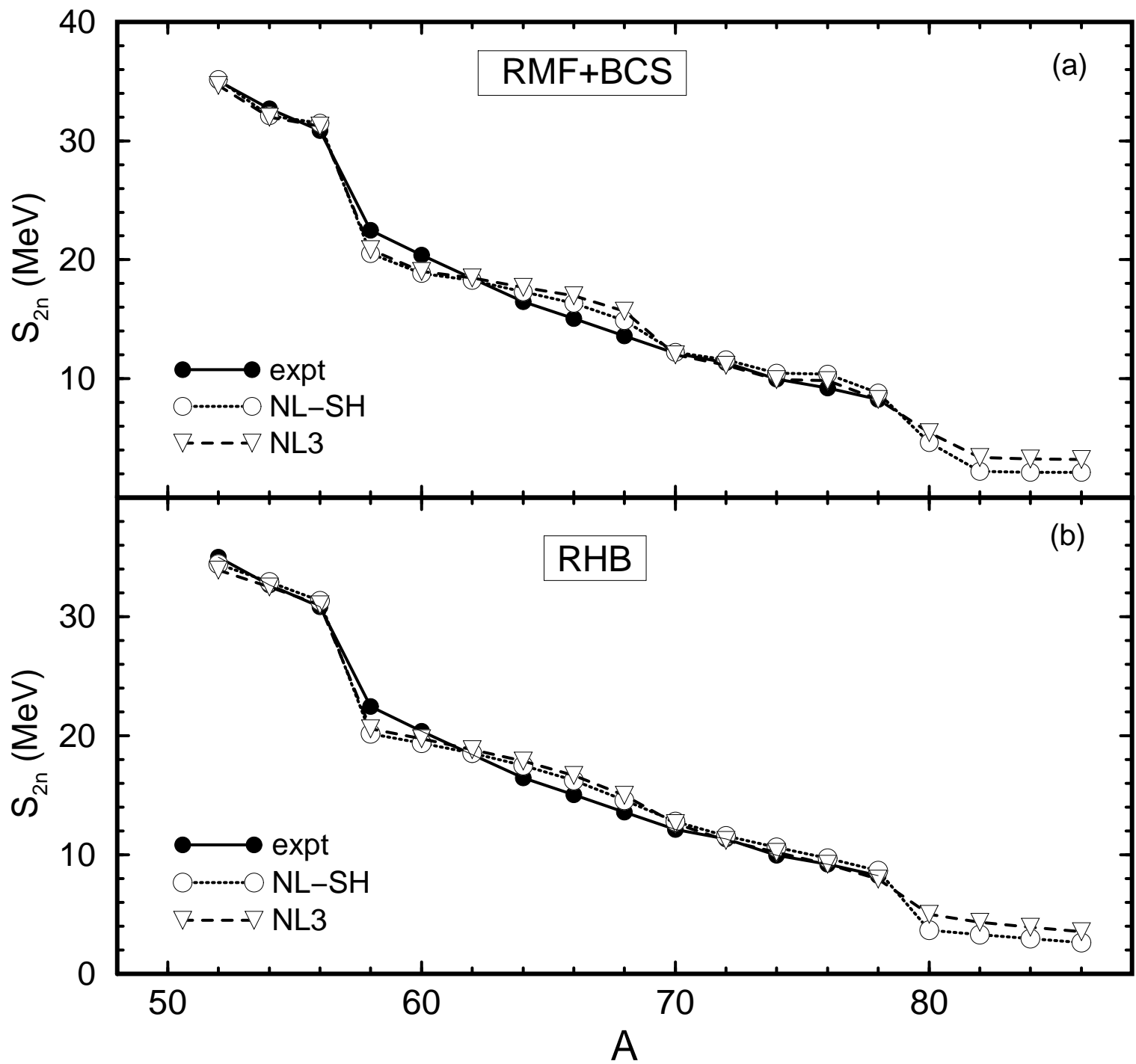
\item[Fig. 4] The $S_{2n}$ values for Ni isotopes in (a) RMF+BCS and 
(b) RHB approach. Within each approach, a comparison of the results of 
the forces NL-SH and NL3 is made with the experimental data.

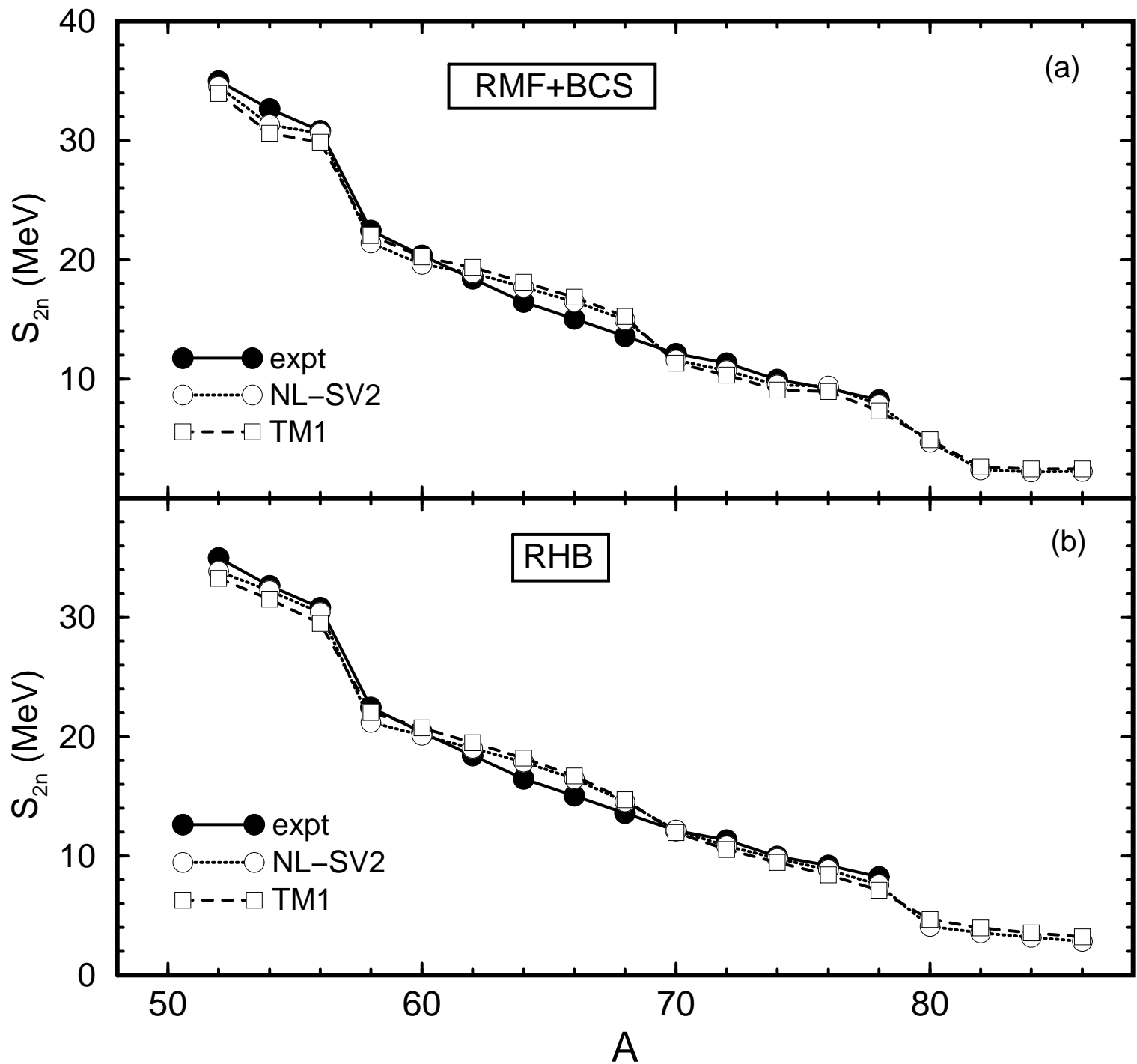
\item[Fig. 5] The $S_{2n}$ values for Ni isotopes in (a) RMF+BCS and 
(b) RHB approach with the forces with the non-linear scalar and vector 
self-coupling. A comparison of the results of the forces NL-SV2 and TM1 
is made with the experimental data.

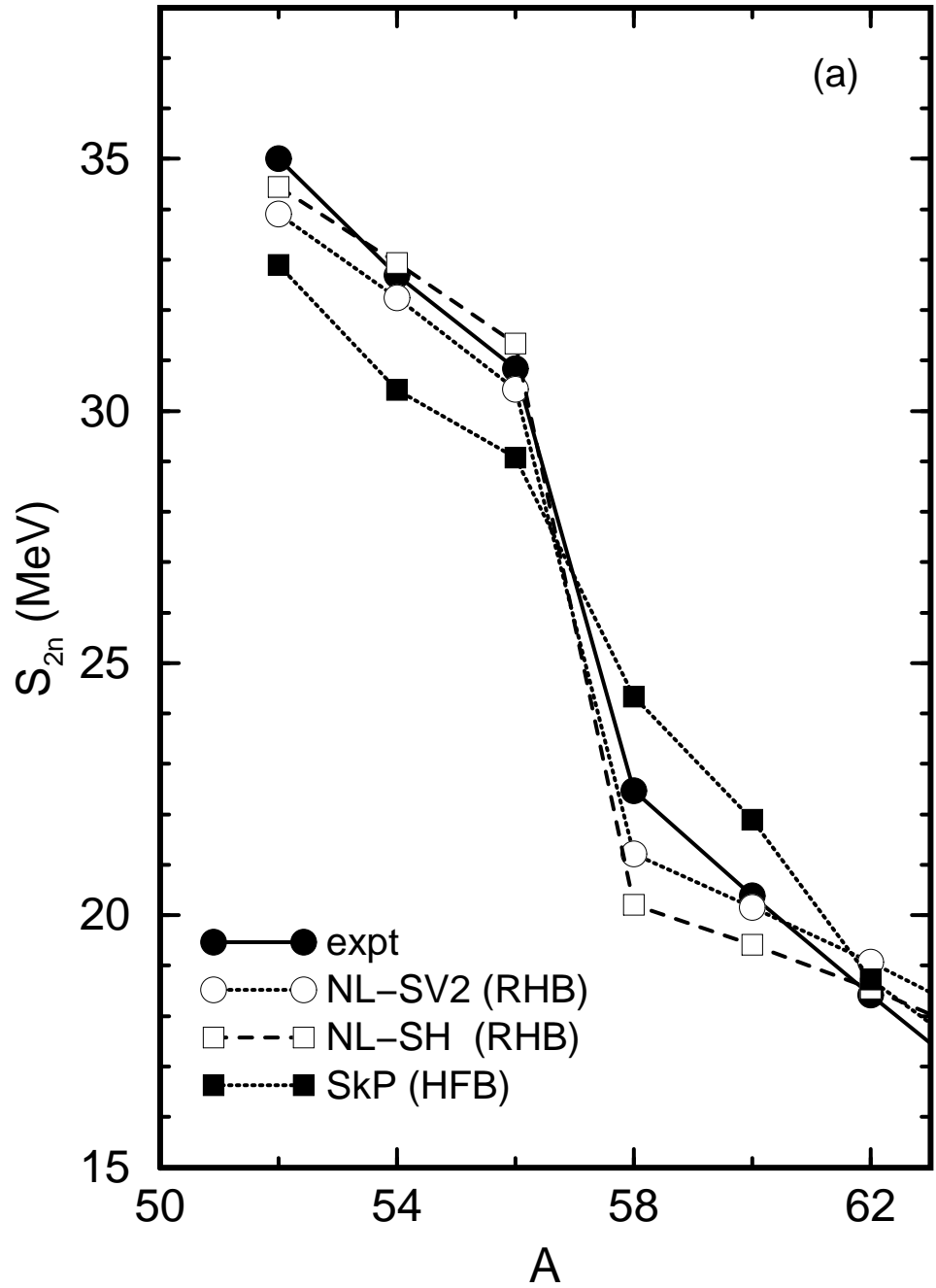
\item[Fig. 6]  A comparison of the 2-neutron separation energy
  $S_{2n}$ of Ni isotopes using the force NL-SH with the scalar 
self-coupling and using the force NL-SV2 with the scalar-vector 
self-coupling in RHB. The results are shown in the region of closed 
neutron shell (a) N=28 (A=56) and (b) N=50 (A=78). A comparison with 
the experimental data is made. The results from the HFB approach with 
the Skyrme force SkP are also shown for comparison \cite{doba_pri.94}.

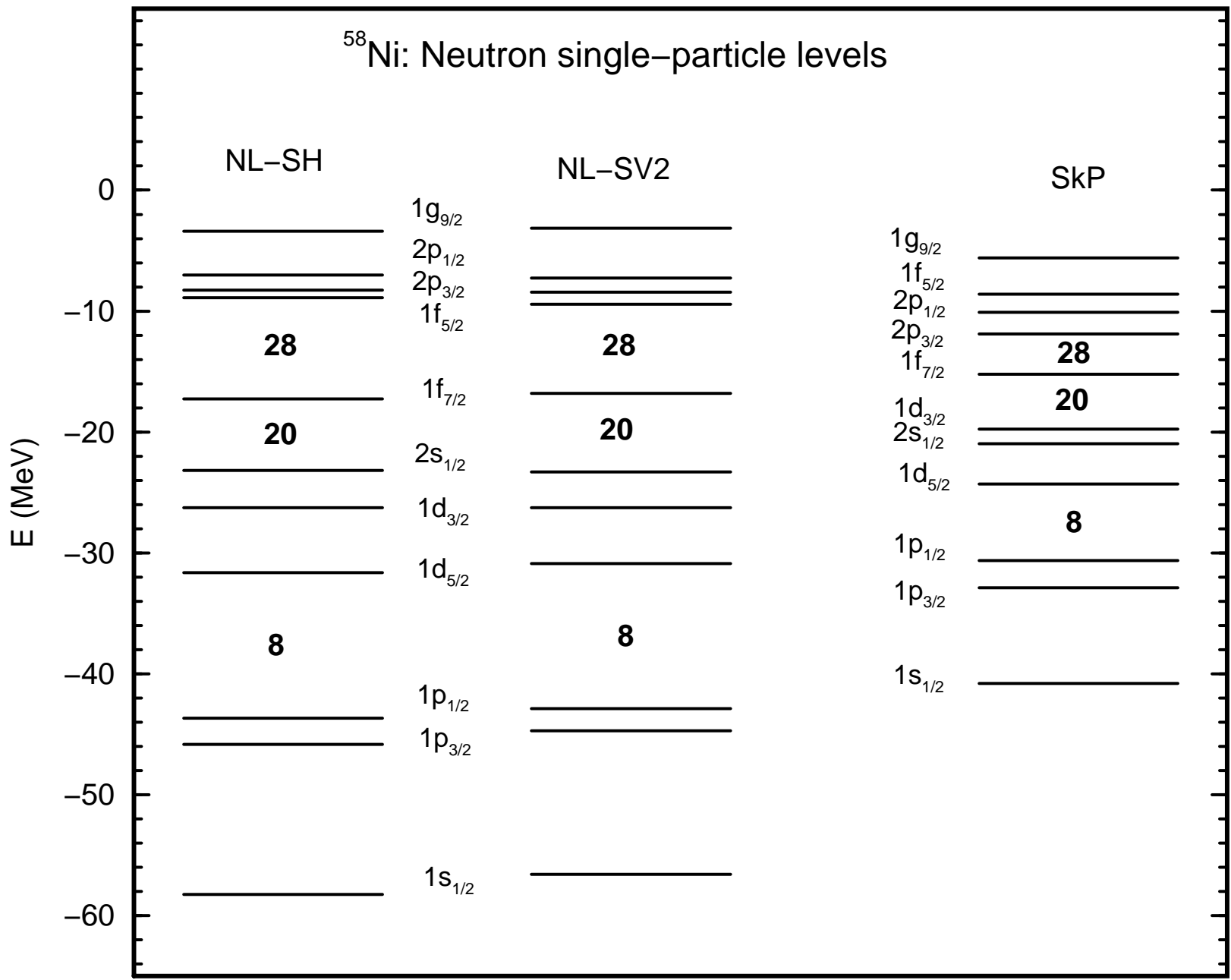
\item[Fig. 7] The neutron single-particle spectrum for the nucleus
$^{58}$Ni calculated in the canonical basis in RHB with the forces
NL-SH and NL-SV2. A reduction in the shell gap at N=28 with the force
NL-SV2 can be seen, as compared to NL-SH. The corresponding spectrum
for SkP shows a drastic reduction in the N=28 shell gap.

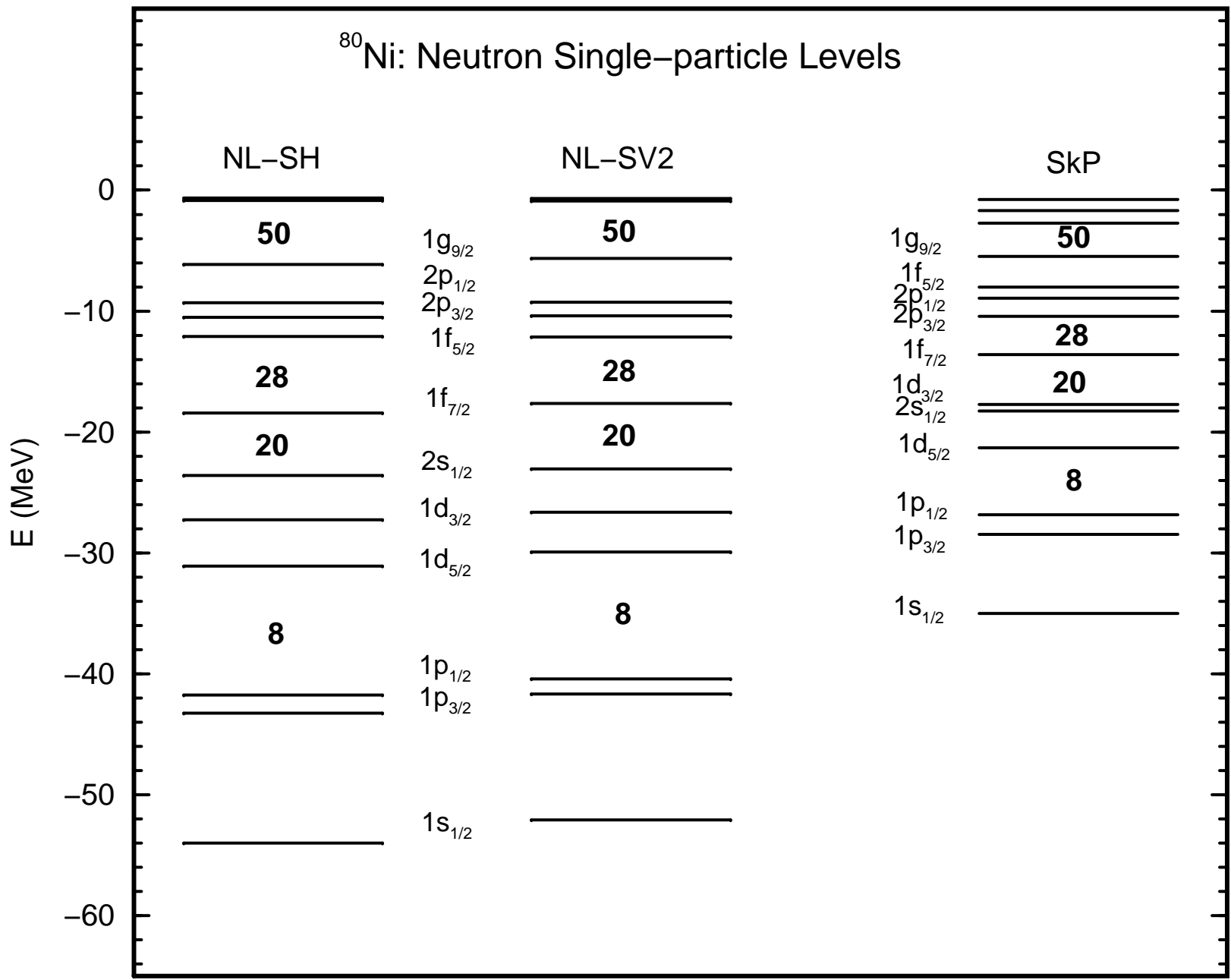
\item[Fig. 8] The same as Fig. 7, for the nucleus $^{80}$Ni. The
shell gap at N=50 with the force SkP is significantly reduced
as compared to that with NL-SH and NL-SV2.

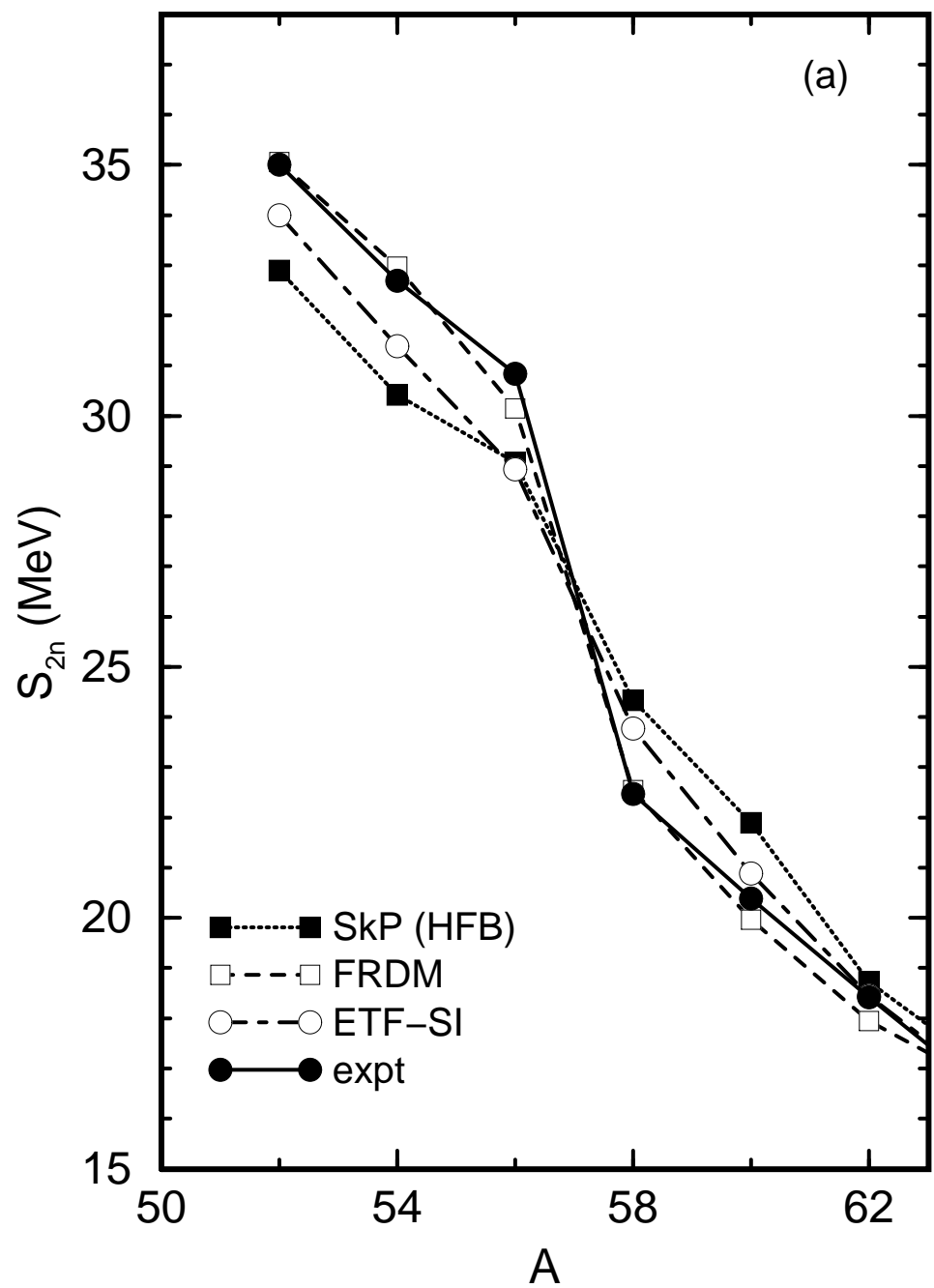
\item[Fig. 9] The $S_{2n}$ values in the region (a) N=28 and (b) N=50
from the mass models FRDM, ETF-SI and the HFB approach with SkP.
A comparison is made with the experimental data available.

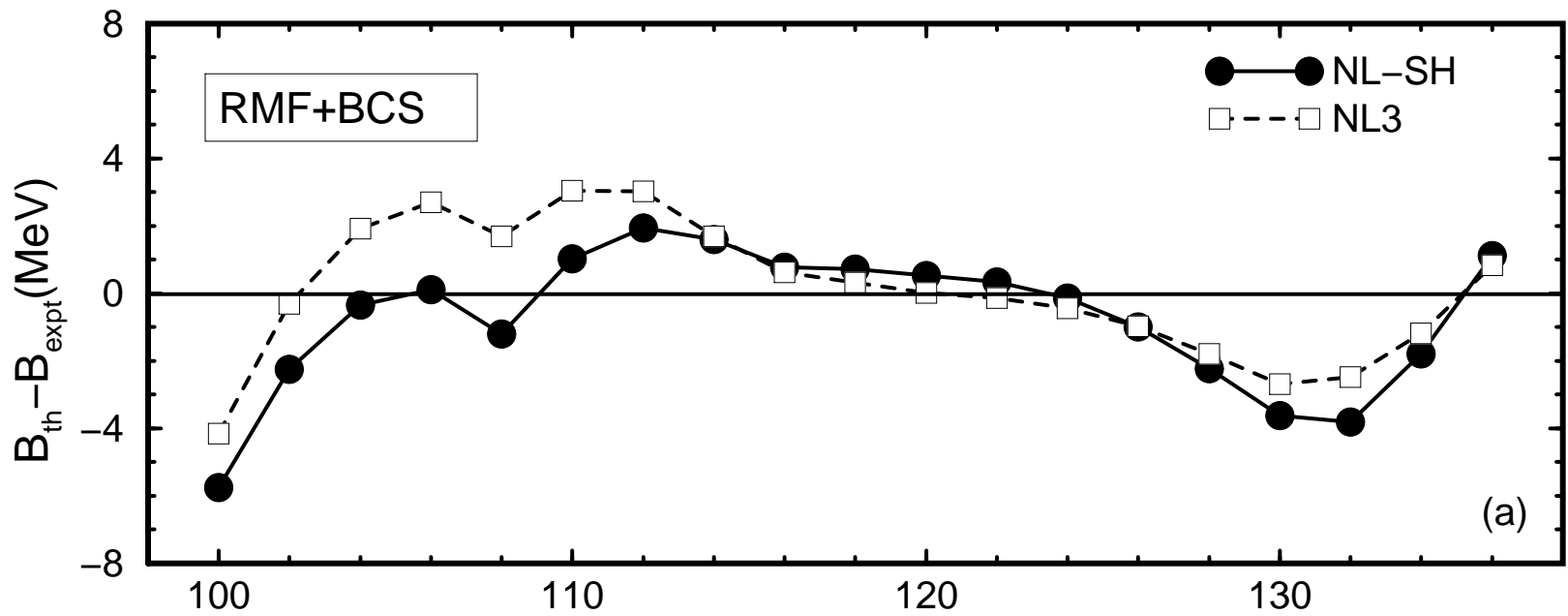
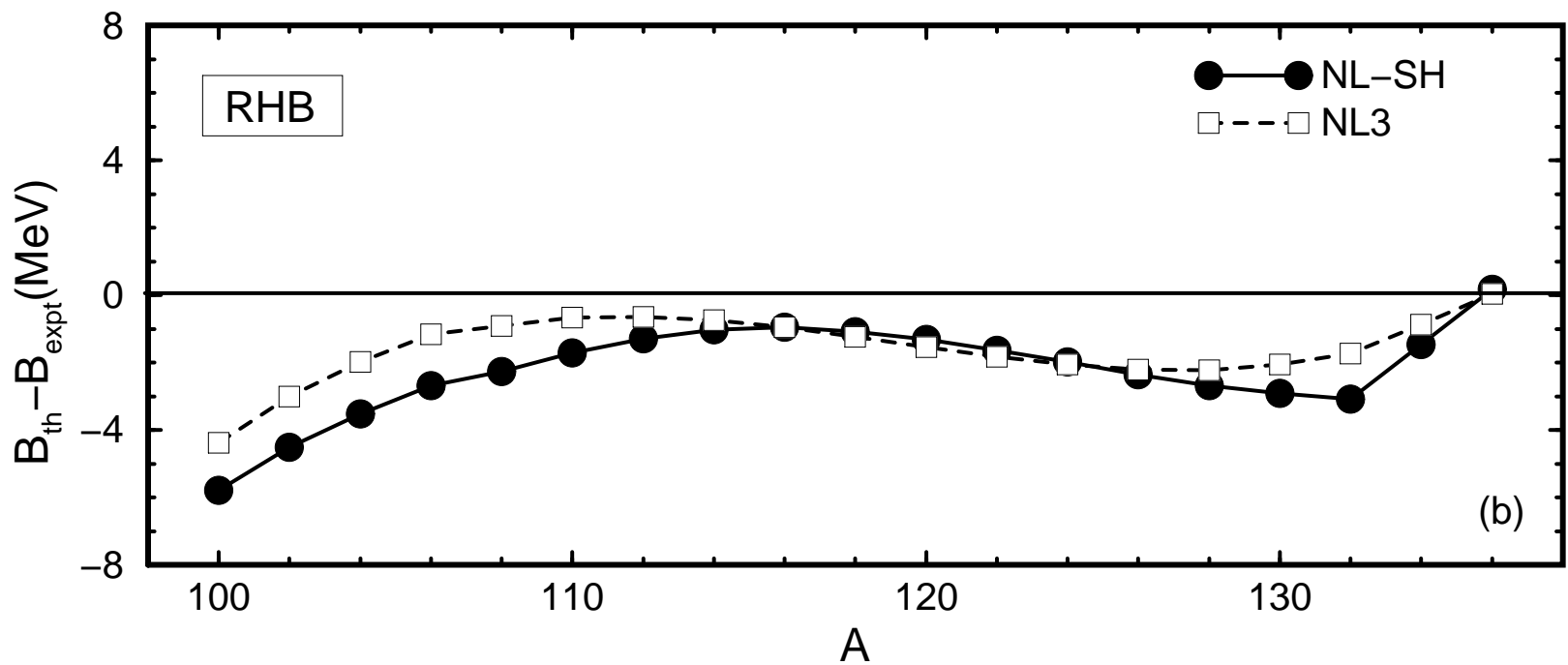
\item[Fig. 10] The binding energies of Sn isotopes in the (a) RMF+BCS
and (b) RHB approach with the forces NL-SH and NL3.

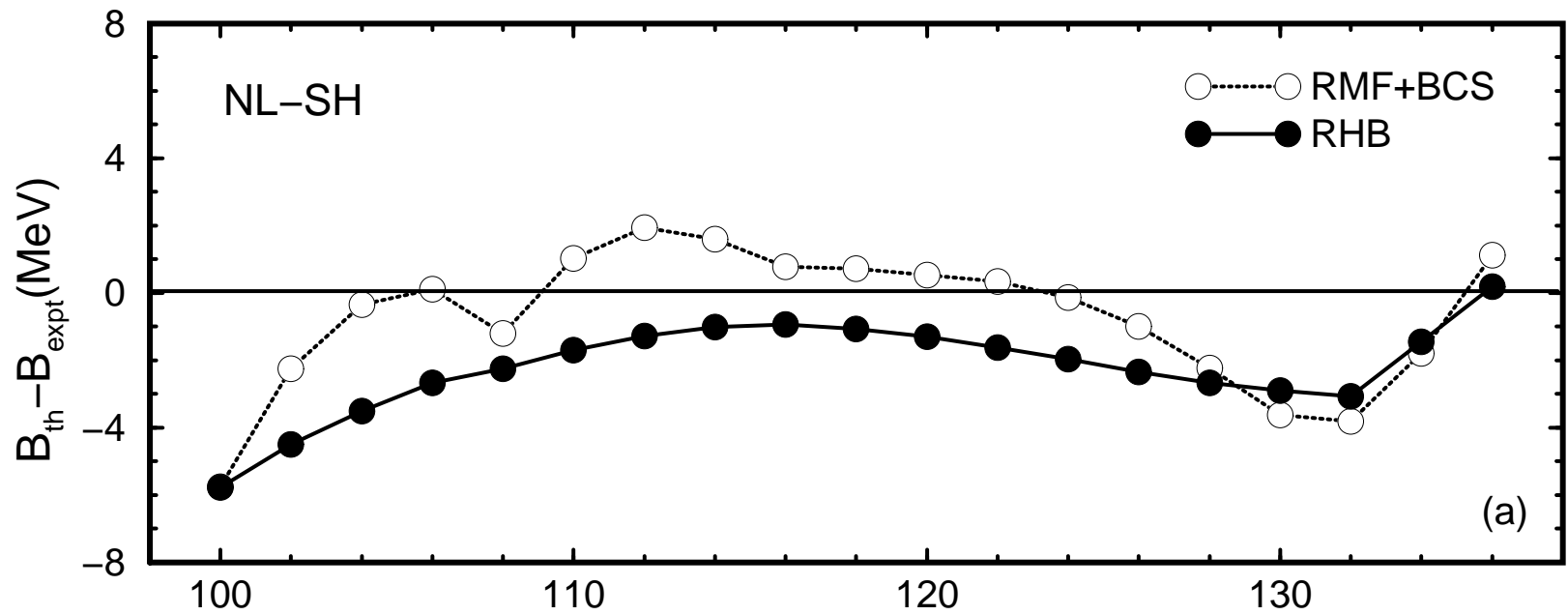
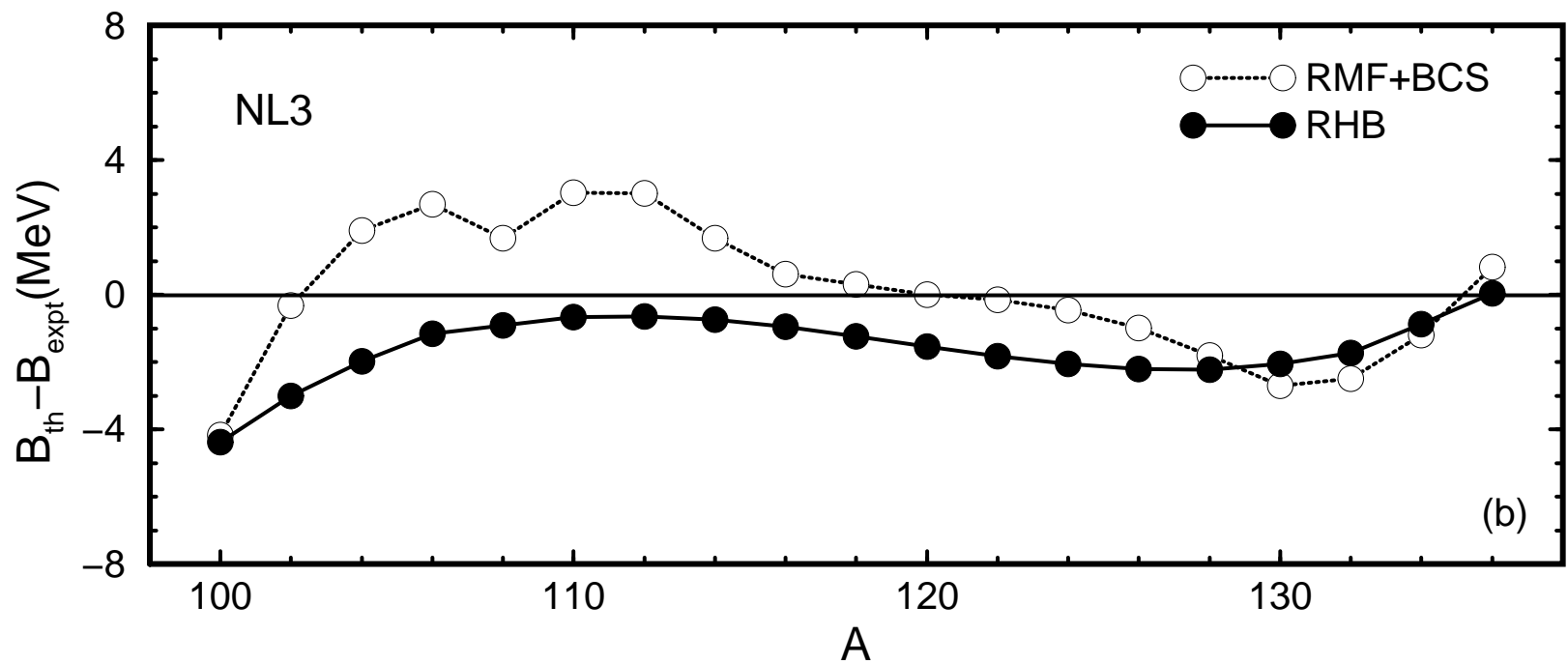
\item[Fig. 11] A comparison of the binding energies of Sn nuclei in
the RMF+BCS and RHB approach with the forces (a) NL-SH and (b) NL3.
A smoothening of the energies in the RHB is observed.

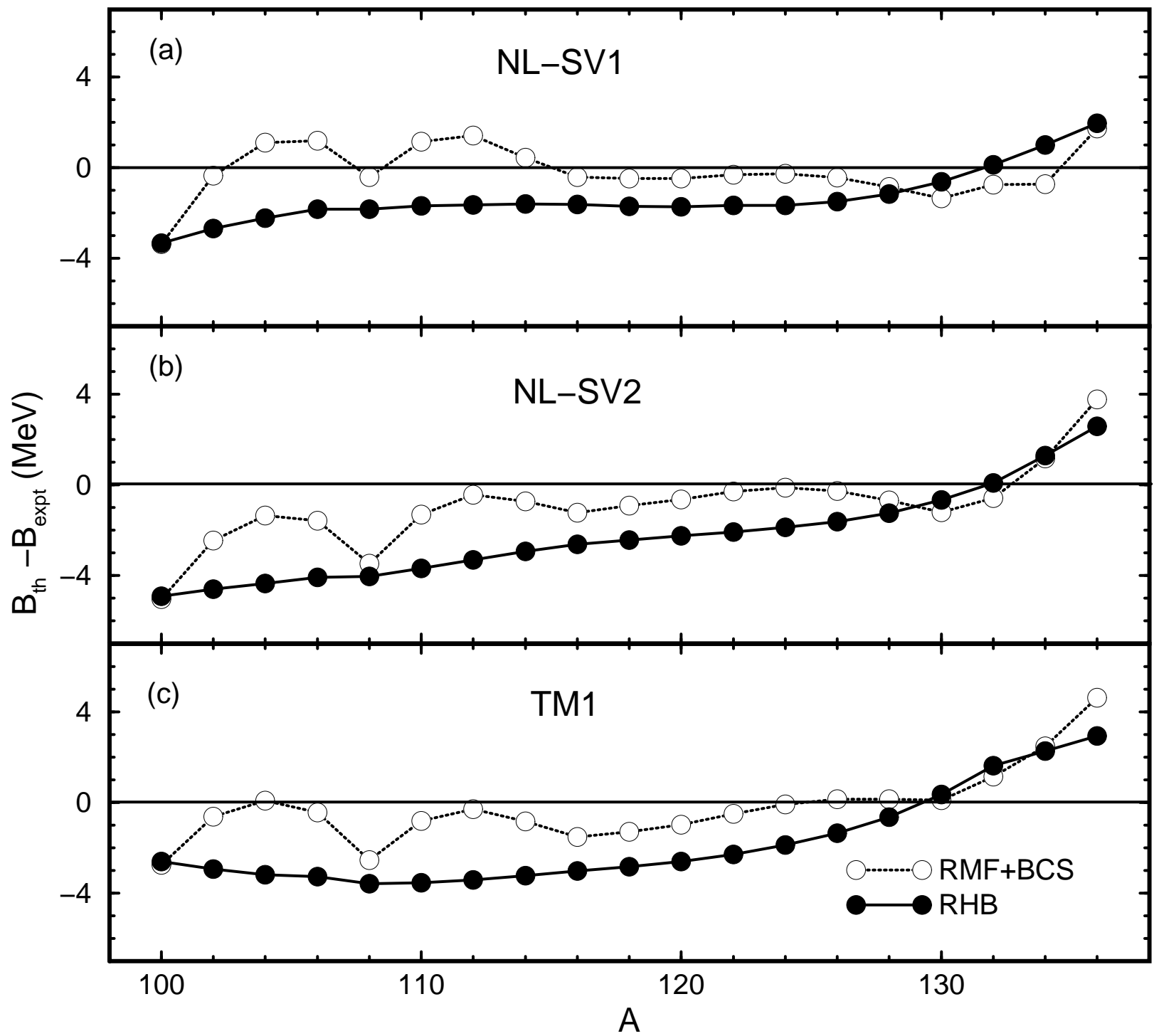
\item[Fig. 12] The binding energies of Sn nuclei using the forces
with the vector self-coupling of $\omega$ meson (a) NL-SV1,
(b) NL-SV2 and (c) TM1, compared with the experimental values,
in RMF+BCS and RHB approach.

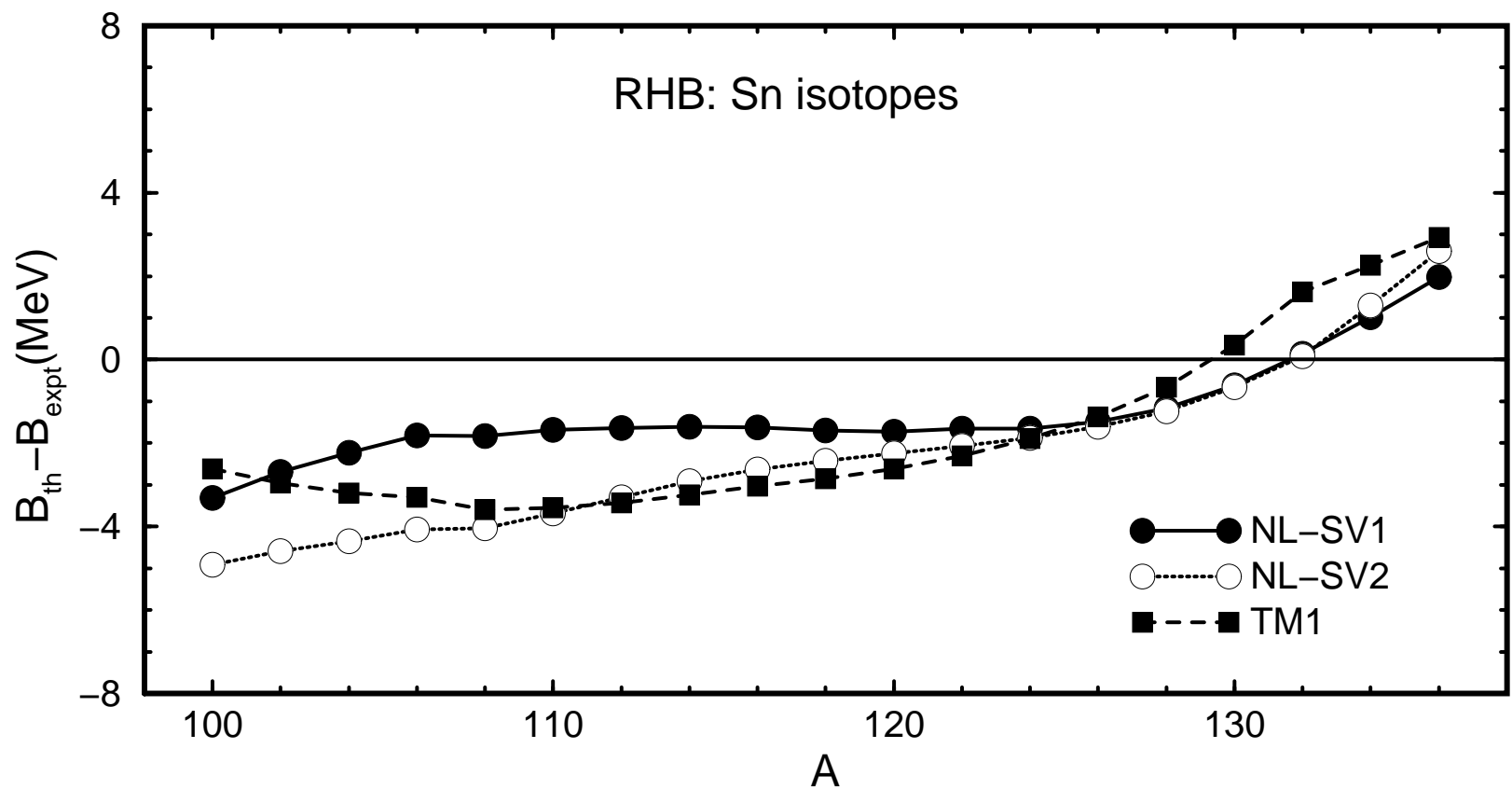
\item[Fig. 13] The binding energies of Sn isotopes obtained with
the RHB approach with the self-consistent pairing. A comparison
is made between the results from the forces NL-SV1, NL-SV2 and TM1.

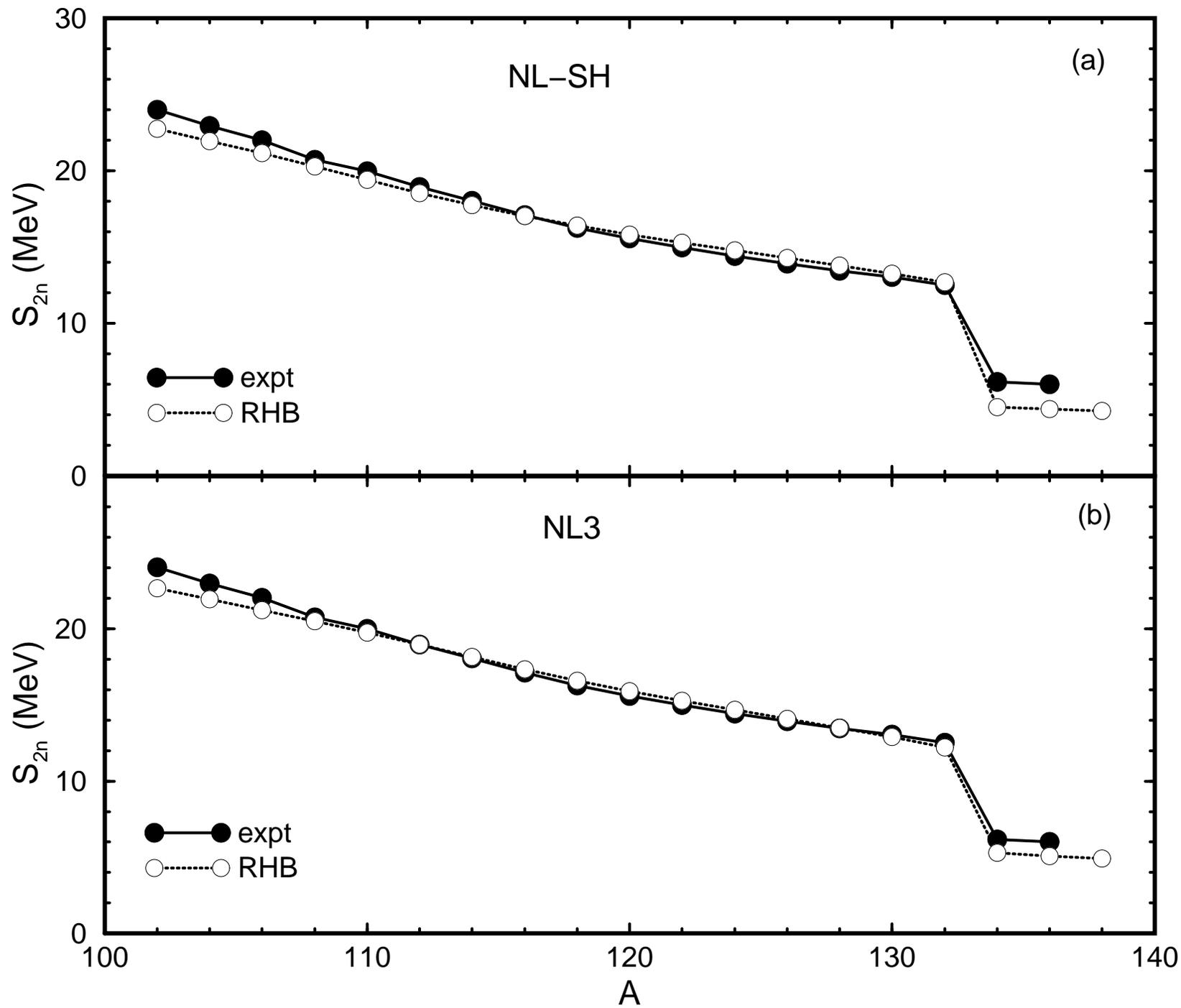
\item[Fig. 14] The $S_{2n}$ values for Sn isotopes obtained in the RHB 
approach with the forces (a) NL-SH and (b) NL3. The experimental data
about A=132 shows a strong kink indicating strong shell effects about
N=82.

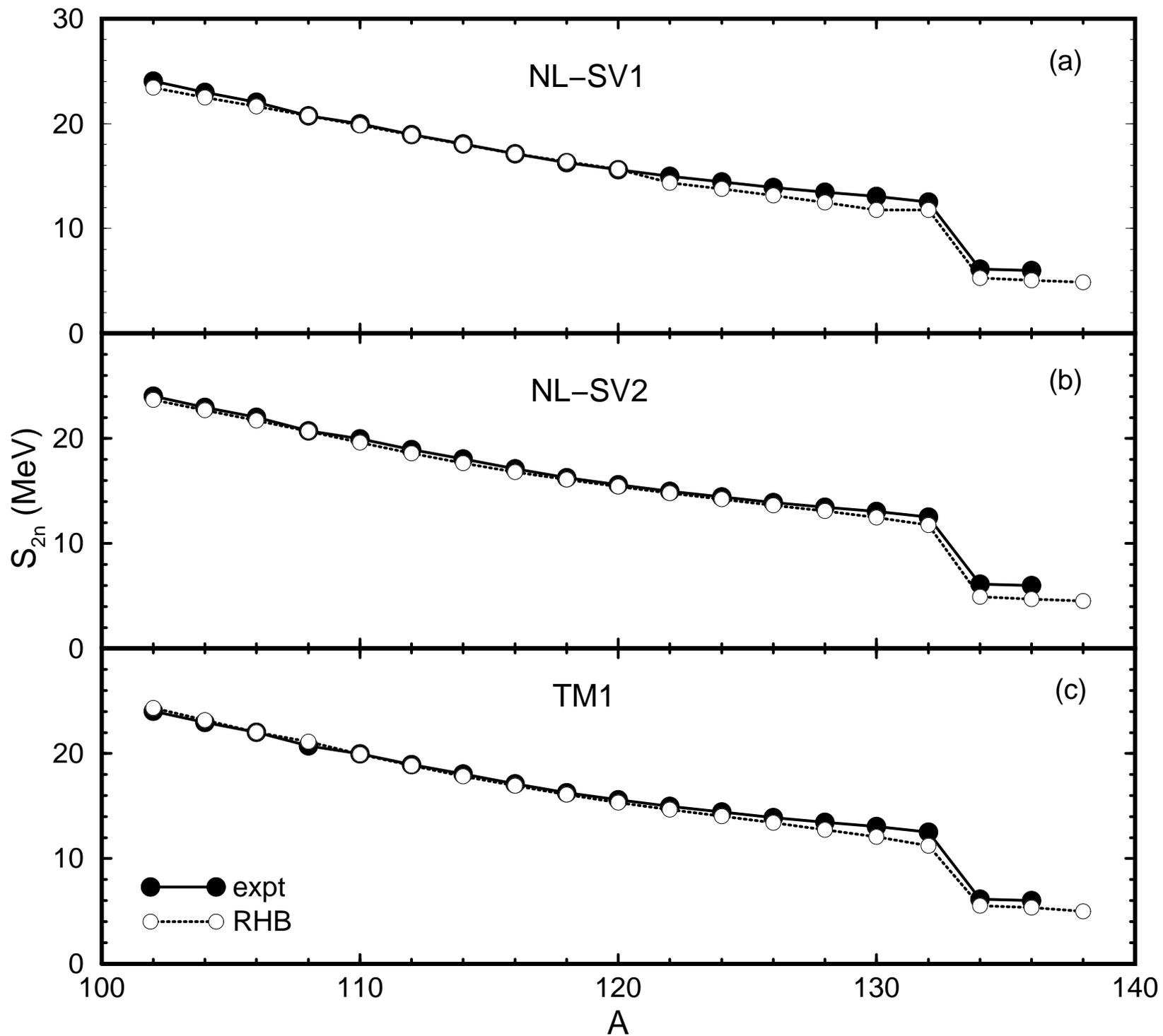
\item[Fig. 15] The $S_{2n}$ values for Sn isotopes obtained in the
RHB approach with the forces (a) NL-SV1, (b) NL-SV2 and (c) TM1.
The experimenetal data are shown for the comparison.

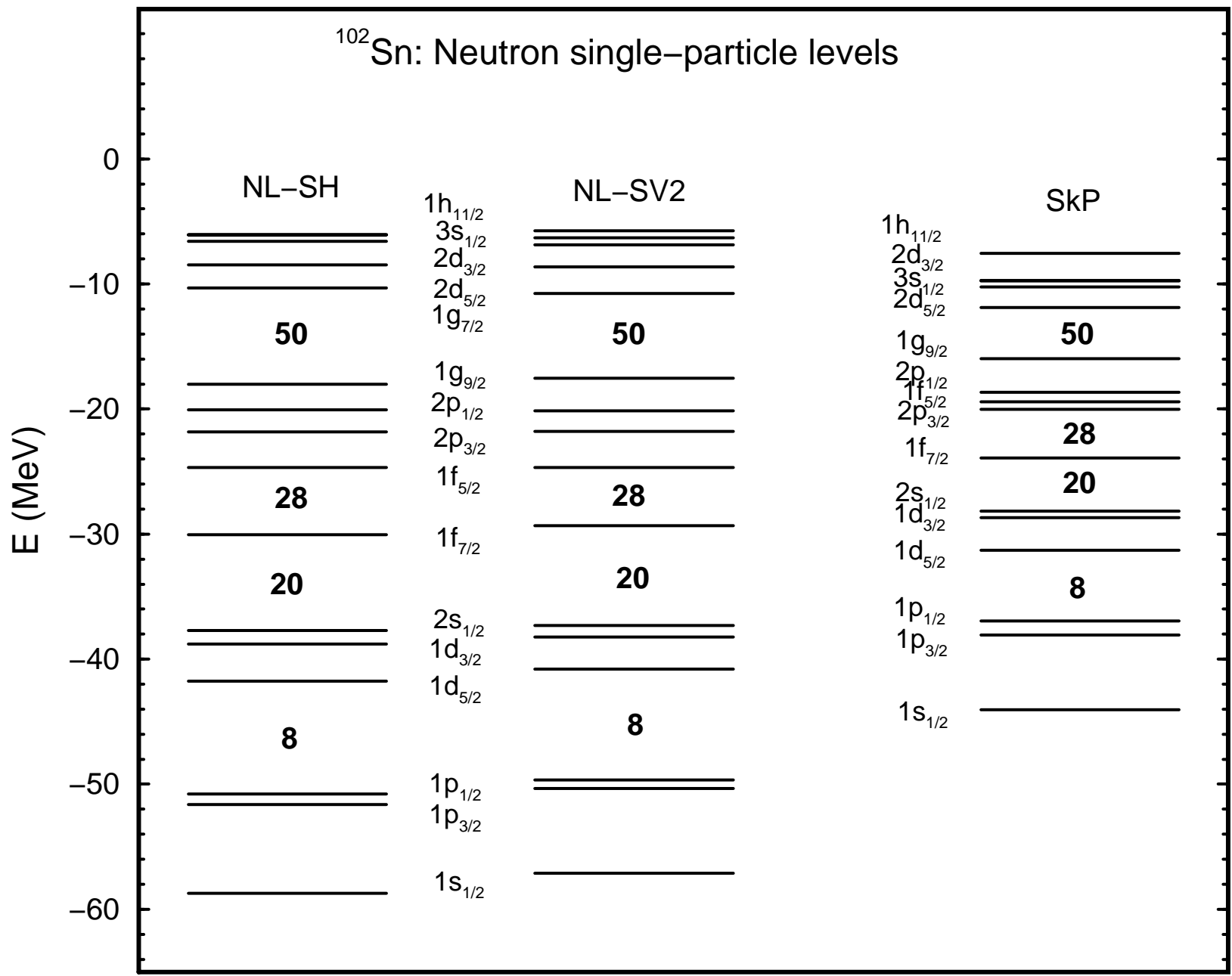
\item[Fig. 16] The neutron single-particle spectrum of $^{102}$Sn
obtained in the canonical basis in the RHB approach with the forces
NL-SH and NL-SV2. The levels with the force SkP are also shown for
comparison. 

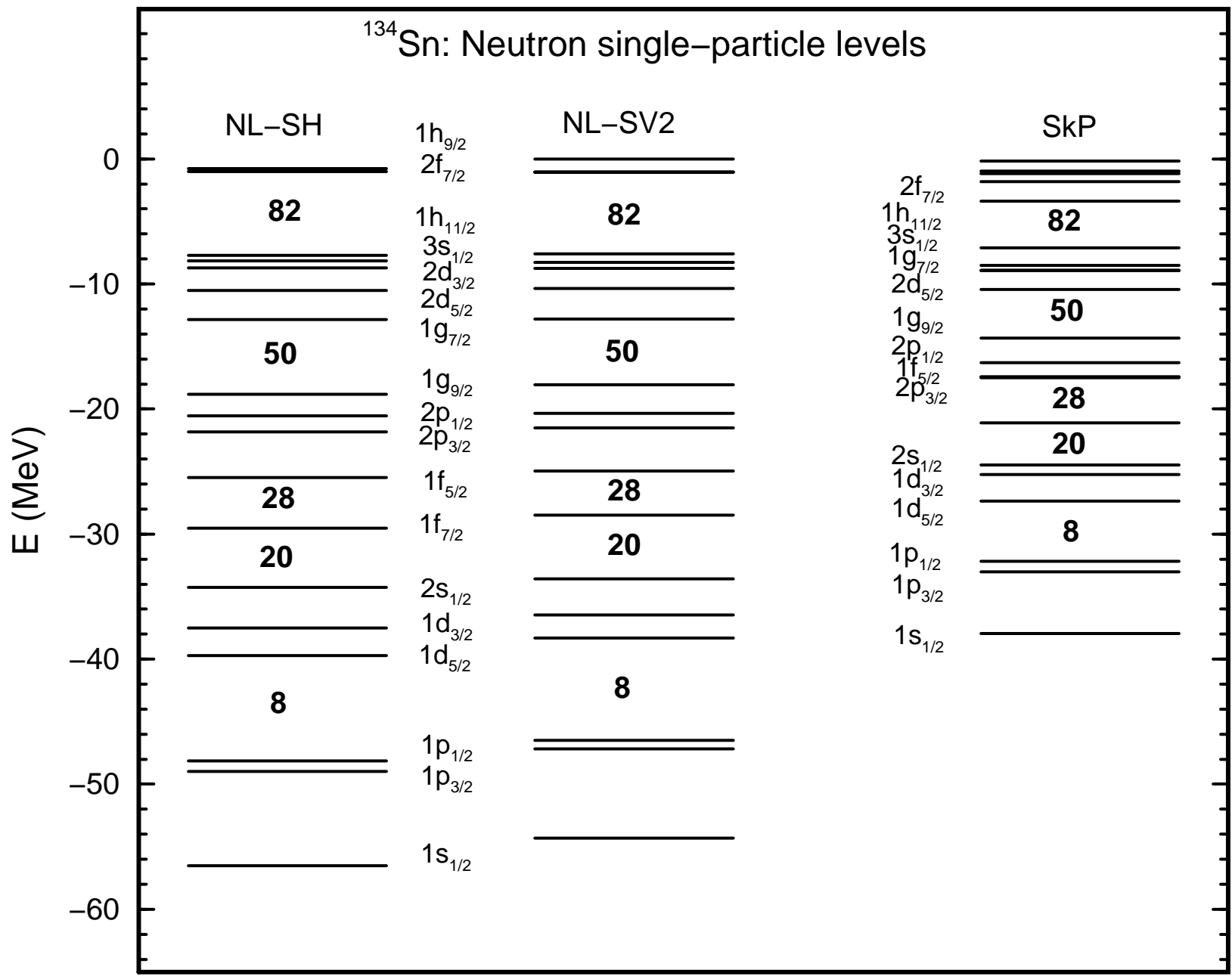
\item[Fig. 17] The same as Fig. 16, for $^{134}$Sn. 

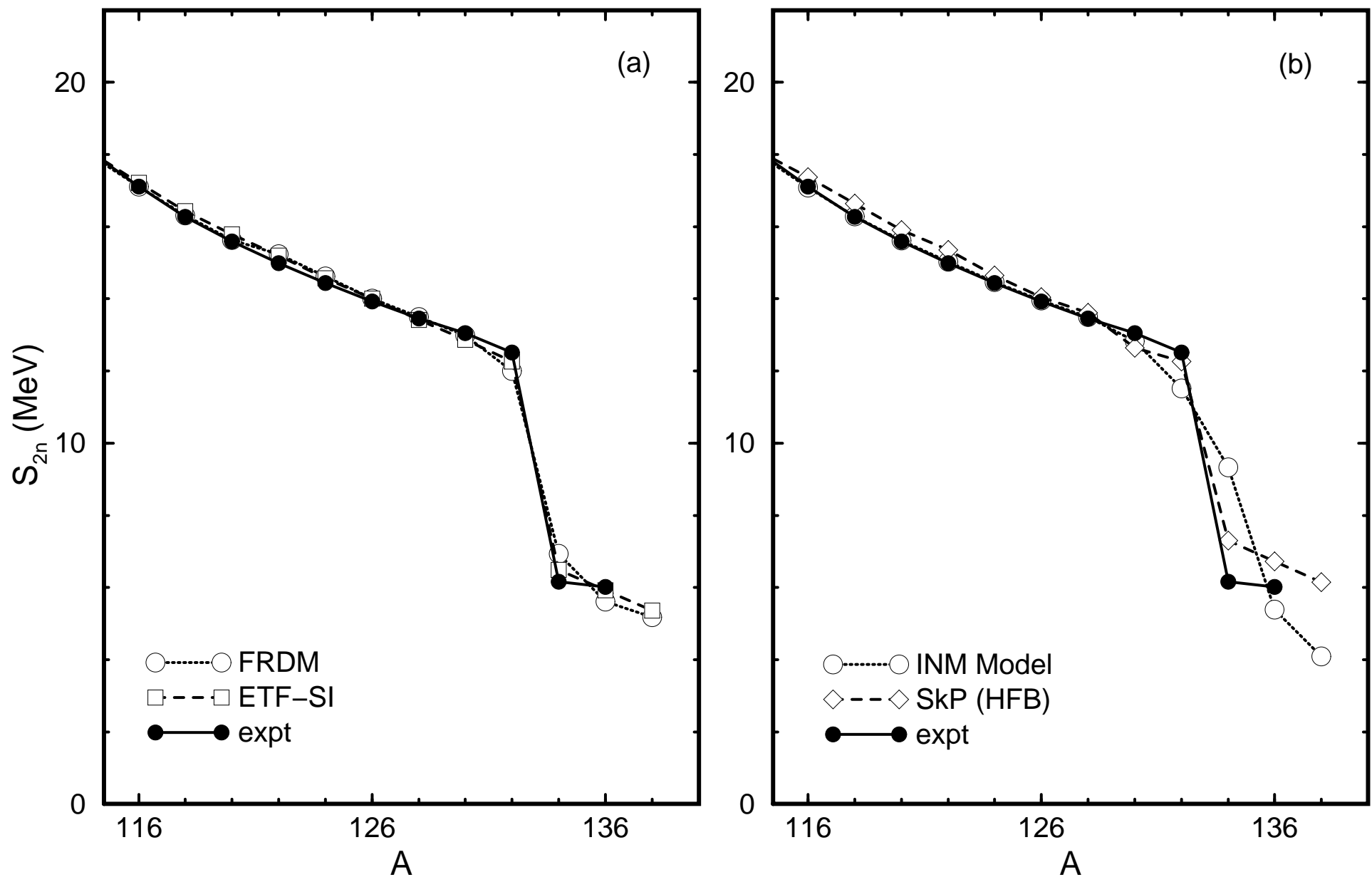
\item[Fig. 18] A comparative view of the $S_{2n}$ values for Sn isotopes
from the mass models FRDM \cite{FRDM.95}, ETF-SI \cite{Abou.95},
and INM model \cite {Satpathy.99} in the region about N=82. 

\end{description}
\vfill

\newpage
\noindent
\begin{table}
\vspace{-0.5cm}
\begin{center}
\caption{\sf The Lagrangian parameters of the forces with the
non-linear scalar self-coupling NL-SH, NL3, and the forces with the
scalar-vector self-coupling NL-SV1, NL-SV2 
and TM1 used in the calculations.}
\bigskip
\begin{tabular}{l l l l l l l l l l }
\hline\\
&                  & & NL-SH     &  NL3     &&&  NL-SV1     &  NL-SV2  & TM1       \\
\\
\hline
&M                 & & 939.0     & 939.0    &&&  939.0      &  939.0    &  938.0    \\
&$m_{\sigma}$      & & 526.05921 & 508.1941 &&&  510.03488  &  519.81202 & 511.198  \\
&$m_{\omega}$      & & 783.0     & 782.5010 &&&  783.       &  783.0     & 783.0    \\
&$m_{\rho}$        & & 763.0     & 763.0    &&&  763.       &  763.0     & 770.0   \\
&$g_{\sigma}$      & & 10.44355  &  10.2169 &&& 10.12479    &  10.32001 & 10.0289 \\
&$g_{\omega}$      & &12.9451    &  12.8675 &&& 12.72661    &  12.88233 & 12.6139 \\
&$g_{\rho}$        & &4.38281    &   4.4744 &&&  4.49197    &  ~4.50144 & 4.6322  \\
&$g_{2}$           & &-6.90992   & -10.4307 &&& -9.24058    &  -6.86061 & -7.2325 \\
&$g_{3}$           & &-15.83373  & -28.8851 &&& -15.388      &  0.36754 & 0.6183  \\
&$g_{4}$           & &0.0        &  0.0     &&& 41.01023    &  72.38965 & 71.30750 \\
\hline
\end{tabular}
\end{center}
\end{table}
\vfill

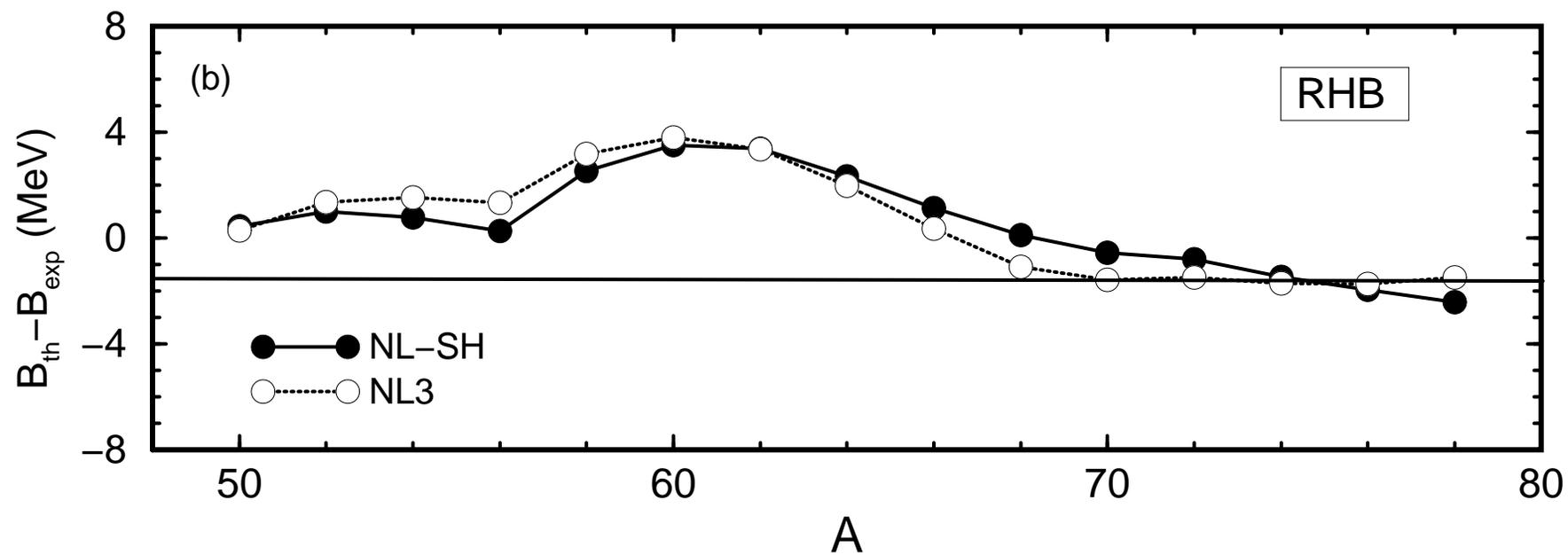
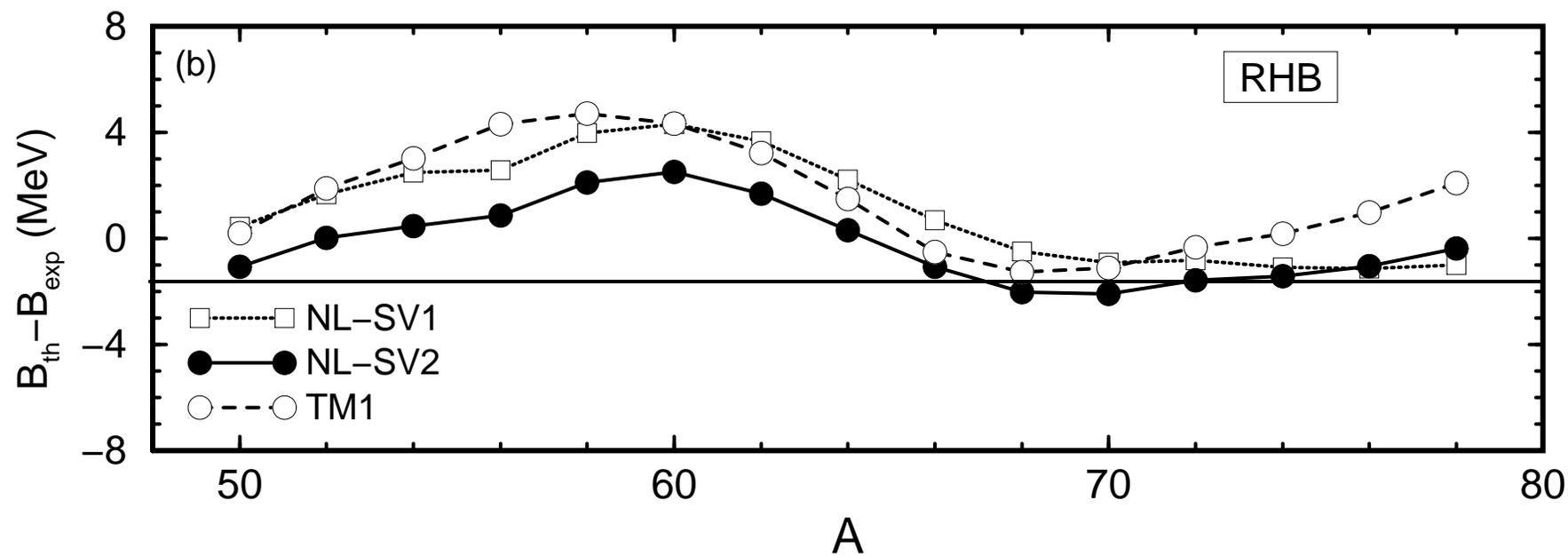
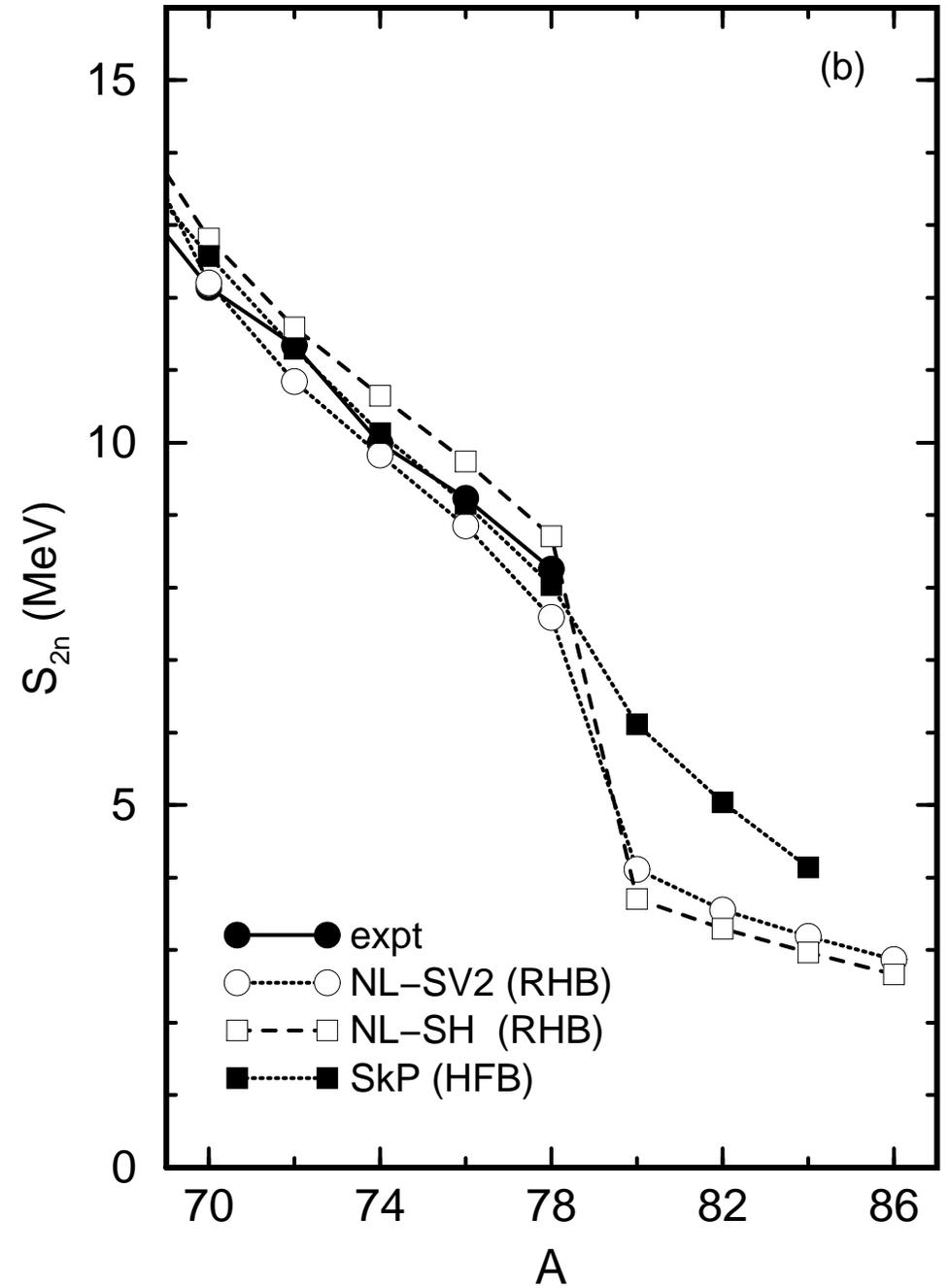
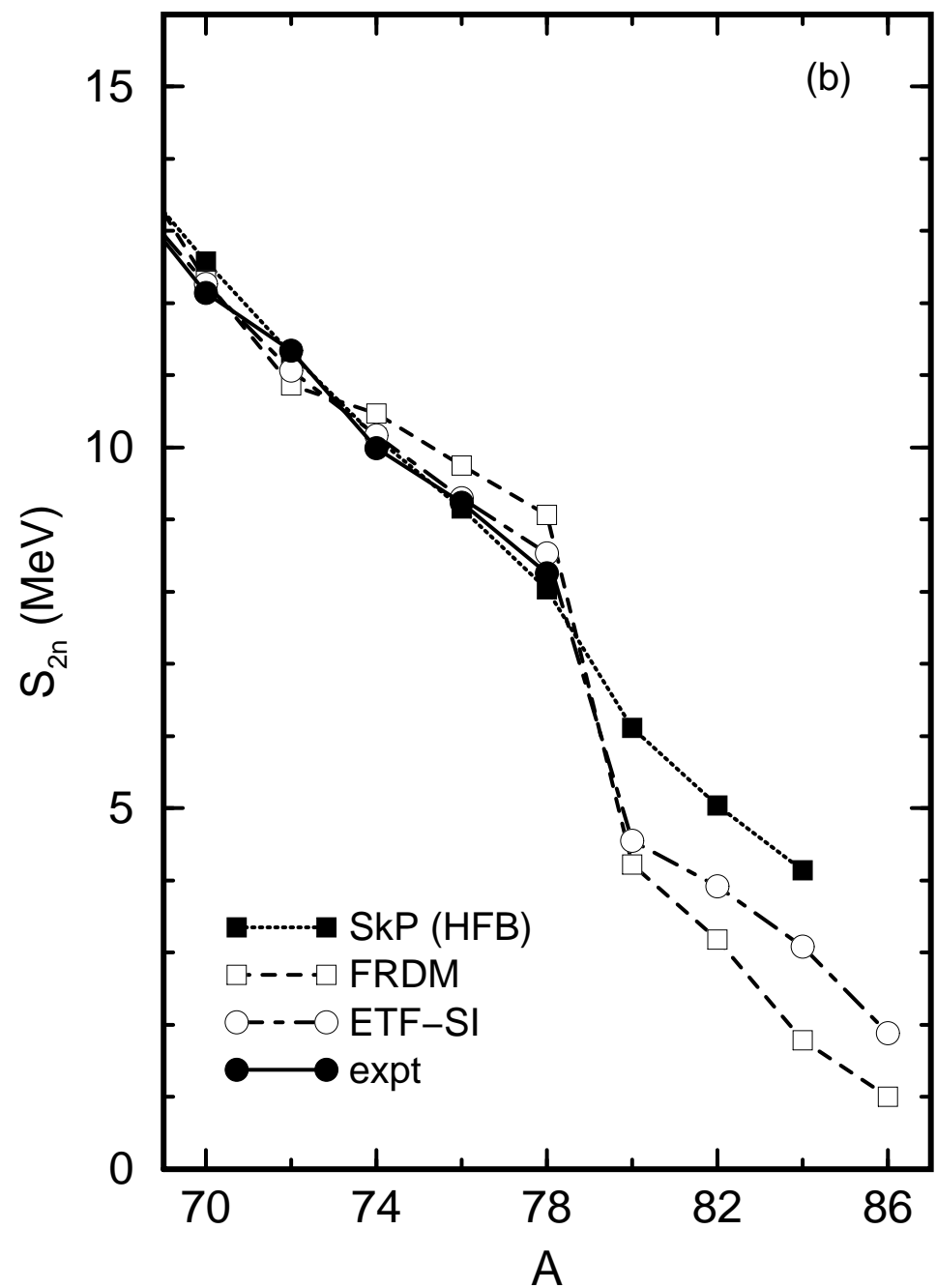
\end{document}